\documentclass[reprint,aps,prl,superscriptaddress,amsmath,amssymb]{revtex4-1}

\usepackage{graphicx}
\usepackage{xcolor}
\usepackage{braket} 
\usepackage{bbold}
\usepackage{changepage}
\usepackage{setspace}
\usepackage[colorlinks=true,urlcolor=blue,citecolor=blue,linkcolor=blue,bookmarks=false,pdfstartview={FitH}]{hyperref}
\newcommand{\be}{\begin{equation}}
\newcommand{\ee}{\end{equation}}
\newcommand{\bea}{\begin{eqnarray}}
\newcommand{\eea}{\end{eqnarray}}

\newcommand{\la}{\left\langle}
\newcommand{\ra}{\right\rangle}
\newcommand{\lb}{\left[}
\newcommand{\rb}{\right]}

\renewcommand{\Re}{{\rm \, Re\,}}
\renewcommand{\Im}{{\rm \, Im\,}}
\renewcommand{\vec}[1]{{\bf #1}}
\def\nn{\nonumber\\}

\begin{document}
    \title{Observation of Chiral Surface Excitons in a Topological Insulator Bi$_2$Se$_3$}

\author{H.-H.~Kung}
\email{sean.kung@ubc.ca}
\altaffiliation{Present address: Quantum Matter Institute, University of British Columbia, Vancouver, BC V6T1Z4, Canada}
\affiliation{Department of Physics \& Astronomy, Rutgers University, Piscataway, New Jersey 08854, USA}
\author{A. P.~Goyal}
\affiliation{Department of Physics, University of Florida, Gainesville, Florida 32611, USA}
\author{D. L.~Maslov}
\email{maslov@phys.ufl.edu}
\affiliation{Department of Physics, University of Florida, Gainesville, Florida 32611, USA}
\author{X.~Wang}
\affiliation{Department of Physics \& Astronomy, Rutgers University, Piscataway, New Jersey 08854, USA}
\affiliation{Rutgers Center for Emergent Materials, Rutgers University, Piscataway, New Jersey 08854, USA}
\author{A.~Lee}
\affiliation{Department of Physics \& Astronomy, Rutgers University, Piscataway, New Jersey 08854, USA}
\author{A. F.~Kemper}
\affiliation{Department of Physics, North Carolina State University, Raleigh, North Carolina 27695, USA}
\author{S.-W.~Cheong}
\affiliation{Department of Physics \& Astronomy, Rutgers University, Piscataway, New Jersey 08854, USA}
\affiliation{Rutgers Center for Emergent Materials, Rutgers University, Piscataway, New Jersey 08854, USA}
\author{G.~Blumberg}
\email{girsh@physics.rutgers.edu}
\affiliation{Department of Physics \& Astronomy, Rutgers University, Piscataway, New Jersey 08854, USA}
\affiliation{National Institute of Chemical Physics and Biophysics, 12618 Tallinn, Estonia}

\begin{abstract}
The protected electron states at the boundaries or on the surfaces of topological insulators (TIs) have been the subject of intense theoretical and experimental investigations. 
Such states are enforced by very strong spin-orbit interaction in solids composed of heavy elements.  
Here, we study the composite particles -- \textit{chiral excitons} -- formed by the Coulomb attraction between electrons and holes residing on the surface of an archetypical three-dimensional topological insulator (TI), Bi$_2$Se$_3$. 
Photoluminescence (PL) emission arising due to recombination of excitons in conventional semiconductors is usually unpolarized because of scattering by phonons and other degrees of freedom during exciton thermalization. 
On the contrary,  we observe almost perfectly polarization-preserving PL emission from chiral excitons. 
We demonstrate that the chiral excitons can be optically oriented with circularly polarized light in a broad range of excitation energies, even when the latter deviate from the (apparent) optical band gap by hundreds of meVs, and that the orientation remains preserved even at room temperature. 
Based on the dependences of the PL spectra on the energy and polarization of incident photons, we propose that chiral excitons are made from massive holes and massless (Dirac) electrons, both with chiral spin textures enforced by strong spin-orbit coupling. 
A theoretical model based on such proposal describes 
quantitatively the experimental observations. 
The optical orientation of composite particles, the chiral excitons, emerges as a general result of strong spin-orbit coupling in a 2D electron system. 
Our findings can potentially expand applications of TIs in photonics and optoelectronics.
\end{abstract}

\maketitle

Spin orbit coupling (SOC) plays a central role in spintronic and optoelectronic applications by allowing optical control of spin excitations and detection with circularly polarized light, in the absence of an external magnetic field~\cite{zutic:2004,Shekhter2005,Mak2016}.
This effect is also known as optical orientation,
where non-equilibrium distribution of spin-polarized quasiparticles are optically created in semiconductors with strong SOC~\cite{Planel1984book,DYakonov:book}.
Detailed information on
spin dynamics can be obtained
by studying 
polarized photoluminescence (PL)~\cite{Parsons1969,Ekimov1970,Zakharchenya1971,Gross1973}.
Typically, the degree of PL polarization in semiconductors
decreases rapidly as the excitation photon energy deviates from the optical band gap 
or with heating~\cite{Bonnot1974PRB,Planel1984book,DYakonov:book}.
Elaborative layer and strain engineering are often required to lift 
spin degeneracy of the bulk bands
to achieve higher degree of PL polarization~\cite{Subashiev1999}.
In contrast, nearly complete PL polarization,
observed recently 
in transition metal dichalcogenide (TMD) monolayers up to room temperature, 
was attributed to 
spin-orbit mediated
coupling between the spin and valley degrees of freedom
~\cite{Mak2012,Xu2014,Jones2013}.
The reduced dimensionality suppresses dielectric screening and restricts the number of scattering channels, resulting in long-lived coherent two-dimensional (2D) excitons~\cite{Qiu2013PRL}.
These results have attracted significant interest due to possible applications and also as an insight into the nature of many body interactions in 2D electronic and photonic systems~\cite{Xu2014,Martin2015PRB,Mak2016}.

\begin{figure*}[ht]
	\includegraphics[width=17.8cm]{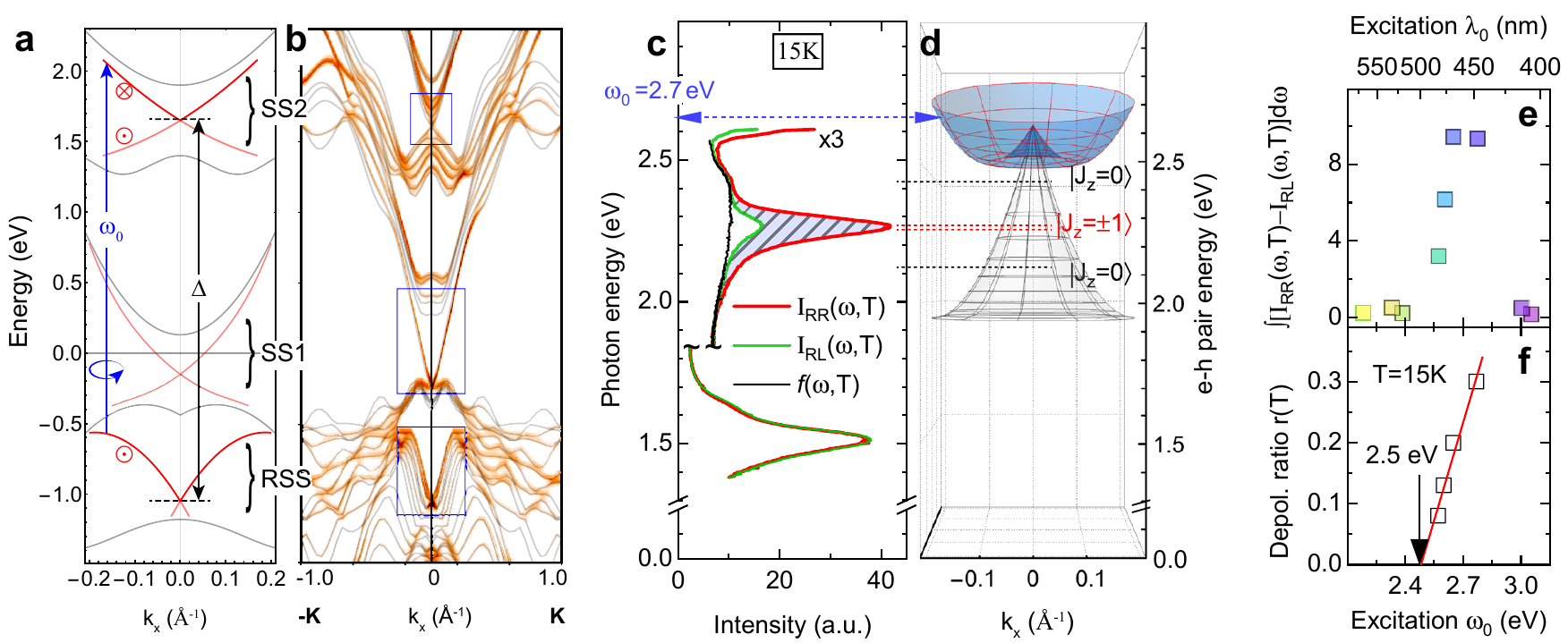}
	\caption{\label{Fig1}
		(a) The electronic band structure near the $\Gamma$ point 
		as inferred from ARPES measurements~\cite{Sobota2013,Soifer2017}.
		The Rashba surface states (RSS) and the unoccupied topological surface states (SS2) are depicted by thick red lines,
		with the in-plane spin orientations denoted by $\odot$ and $\otimes$.
		The low energy surface states (SS1) near the Fermi energy ($E_F$) are 
		depicted by thin red lines. 
		The bulk bands (which do not contribute to circularly polarized PL) and are shown in gray.
		(b) Calculated band structure along the $\Gamma$--K cut in the Brillouin zone of the hexagonal lattice, projected onto the top quintuple layer for $J_z=1/2$. The blue squares highlight the 3 surface bands [See \textit{SI Appendix} Sec.~S2].
		(c) The PL spectra measured with right-circularly polarized 2.7\,eV excitation at 15\,K.
		Right- and left-circularly polarized PL signals are designated by $I_\text{RR}(\omega,T)$ in red, and $I_\text{RL}(\omega,T)$ in green, respectively.
		The black line shows unpolarized PL background, $f(\omega,T)$.
		The intensity in the 1.8--2.6\,eV range is multiplied by factor of 3 for clarity.
		(d) The dispersion relation of non-interacting electron-hole pairs for possible transitions from RSS to SS2, with zero momentum transfer.
		The only transition consistent with the excitation energy employed
		in this study is shown in blue, otherwise shown in gray.
		With finite interaction, the excitonic bound states form below the band minimum, and are denoted by the total angular momenta of the electron-hole pairs, $J_z$.
		(e) The integrated polarized PL intensity, 
		$\int_{2.0}^{2.5\,eV}[I_\text{RR}(\omega,T)-I_\text{RL}(\omega,T)]d\omega$ [shown as the shaded  blue area in (c)], versus excitation energy measured at 15\,K.
		(f) The depolarization ratio \( r(T)\equiv \frac{I_\text{RL}(\omega,T)-f(\omega,T)}{I_\text{RR}(\omega,T)-f(\omega,T)} \) 
		is plotted as a function of ($\omega_0$), with $T\approx 15$\,K.
		The red line is a linear extrapolation to $r(T)=0$, suggesting a minimum excitation threshold energy of 2.5\,eV.}
\end{figure*}

In this paper, we 
discuss a new
class of excitons 
which produce helicity preserving PL.
We use polarization-resolved PL spectroscopy to study the secondary emission from a 2D electronic system
of significant current interest:
the surface state of a three-dimensional (3D) topological insulator (TI), Bi$_2$Se$_3$.
\footnote{Polarized PL was also observed in {Bi}$_{1.95}${In}$_{0.05}${Se}$_3$ crystals [See \textit{SI Appendix}, Sec.~S3].}
In 3D TIs, strong SOC and time-reversal symmetry collaborate to support 
topologically protected
massless 
surface states 
[denoted by SS1 in Figs.~\ref{Fig1}(a) and (b)]
with chiral spin-momentum texture
\cite{Hsieh2009,Zhang2009,Chen2009,Qi2011,Hasan2011}.
Both angular-resolved photoemission (ARPES) data and first-principle calculations show that there are two more surface bands near the Brillouin zone center ($\Gamma$-point) in Bi$_2$Se$_3$~\cite{Sobota2013,Soifer2017,Bugini2017}:
(1) a high-energy unoccupied Dirac cone (SS2) and (2) fully occupied Rashba-like surface states (RSS). 
These bands are depicted by red lines in Fig.~\ref{Fig1}(a) and enclosed in boxes in Fig.~\ref{Fig1} (b).
Due to strong SOC, all the three surface bands 
exhibit spin-momentum locking,
which could lead to optical orientation of 
single-electron spins
and excitons~\cite{Planel1984book}.
So far, most of research on TIs have been focused on 
spin dynamics and collective modes of 
 Dirac fermions in SS1~\cite{Aguilar2012PRL,Wu2016Science,Kung2017PRL,Shao2017NanoL}, and far less is known about the properties of RSS and SS2.
We excite interband transitions between surface states, RSS and SS2 with circularly polarized light and study polarization of PL emission in the backscattering geometry, 
with light being
incident normally on
the crystal surface.

\subsection*{Polarized photoluminescence}
Figure~\ref{Fig1}(c) 
depicts the intensities of right- and left-circularly polarized
PL signals
excited by right-circularly polarized light, 
$I_\text{RR}(\omega,T)$ and 
$I_\text{RL}(\omega,T)$,
 respectively,
where $\omega$ is 
the energy of emitted photons and $T$ is temperature.
Two emission peaks at about $1.5$ and $2.3$\,eV 
in the visible range
behave in a strikingly different ways 
when excited by circularly polarized light. 
Namely, the peak at 1.5\,eV is unpolarized, i.e., emission of right- 
and left-circularly polarized light has same intensity.
The peak at 1.5\,eV 
behaves as an ordinary PL signal observed in conventional semiconductors, where
the memory about the incident photon polarization is lost
during thermalization of optically generated electron-hole pairs.
In contrast, the peak at 2.3\,eV is 
almost fully polarized with the same polarization as the excitation photon, i.e., 
emission occurs
in the RR channel but not in the RL one.

We note that excitons are not usually observed in semimetals and
doped semiconductors, with the
Fermi level crossing the conduction 
band, because the exciton state is likely to be hybridized with the conduction
band states.
Even if the exciton level remains within the gap,
the optically produced electron-hole pairs would 
relax rapidly  to the Fermi energy in a non-radiative way.
In our case,  RSS and SS2 are gapped from the Fermi level, and thus the electron-hole bound state can
decay radiatively, resulting in observed PL.

\begin{figure}[ht]
	\centering
	\includegraphics[width=8.7cm]{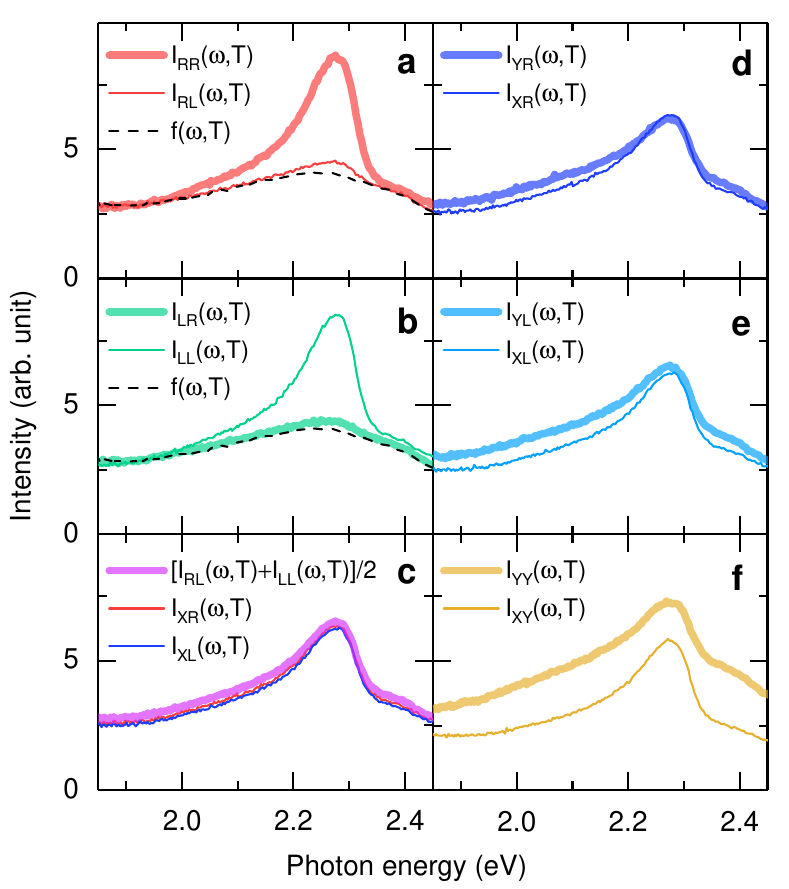}
	\caption{\label{Fig_PolDep}
		Polarization dependence of PL measured with 2.6\,eV excitation at 26\,K.
		(a)--(b), The thick and thin lines show the right- and left-handed PL spectra under right- and left-circularly polarized excitation.
		The black line shows the unpolarized background.
		Panels (c)--(e) compare circularly polarized PL spectra excited with linearly polarized light, where X (Y) denotes linear polarization parallel (orthogonal)
		to the plane of incidence.
		(f), Comparison of the spectra with excitation polarization parallel and orthogonal to the PL polarization.
	}
\end{figure}

To further elucidate the nature of 
polarized PL in Bi$_2$Se$_3$, 
we compare the spectra measured 
in different polarization geometries.
The results are reproducible in the ``time-reversed'' geometry, i.e.,
the polarized PL signals with right- [Fig.~\ref{Fig_PolDep}(a)] and left-circularly polarized excitation [Fig.~\ref{Fig_PolDep}(b)] show the same line shape and intensity.
This suggests that the 
light-emitting
states are doubly degenerate, 
with components 
amenable to independent
excitation
by right- or left circularly-polarized photons.
If relaxation processes of these states preserve 
their
angular momenta, the secondary photons
are emitted with the same polarization as the excitation one.
For reasons that will become clear
later on, we denote these states as $\ket{J_z=1}$ and $\ket{J_z=-1}$ [Fig.~\ref{Fig1}(d)].

In Fig.~\ref{Fig_PolDep}(c)--(f) we show 
intensity of
PL excited with linearly polarized light, with X (Y) denoting linear polarization parallel (orthogonal) to the plane of incidence.
We find that the PL 
signal has almost
the same intensity and line shape in both circular polarization channels, regardless 
of whether
the excitation photon is X or Y polarized [Fig.~\ref{Fig_PolDep}(c)--(e)].
This suggests that a linearly polarized photon, being decomposed into right- and left-circularly polarized ones, can independently excite both the $\ket{J_z=1}$ and $\ket{J_z=-1}$ states.
That  linear polarization 
 is not preserved in the PL process [Fig.~\ref{Fig_PolDep}(f)], and 
 that $I_\text{XL}(\omega,T)$ coincides with 
 $[I_\text{RL}(\omega,T)+I_\text{LL}(\omega,T)]/2$ 
 [Fig.~\ref{Fig_PolDep}(c)], imply that quantum coherence is not preserved during the relaxation of electron-hole pairs.   
As the result, the $\ket{J_z=1}$ and $\ket{J_z=-1}$ excitonic states act as two independent emitters,
which preserve linear but not circular polarization.
This property of emission from Bi$_2$Se$_3$ surface states is in 
contrast to polarized PL observed in TMD monolayers, where both 
circular and linear polarization are preserved due to valley quantum 
coherence~\cite{Jones2013,Xu2014}.   

\subsection*{Dependence on the energy of incident photons}

\begin{figure}[t]
	\centering
	\includegraphics[width=7cm]{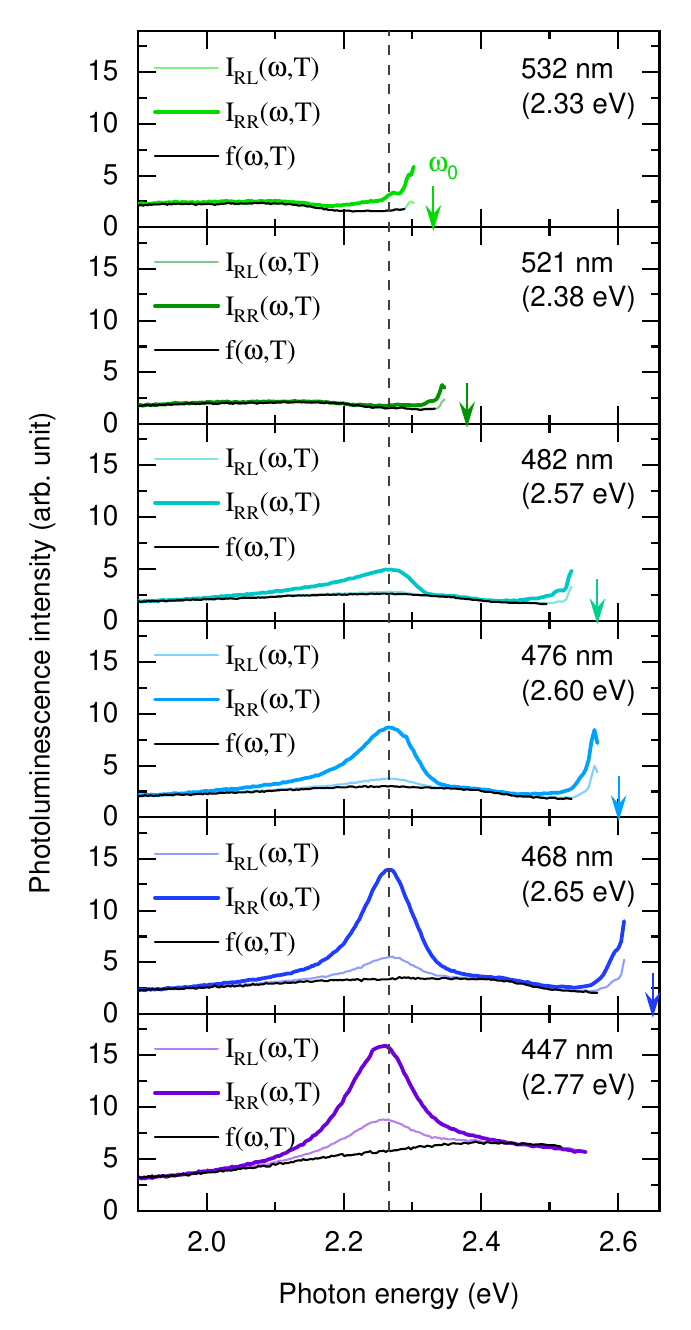}
	\caption{\label{Fig_ExDep}
		Low temperature PL intensity is plotted against photon energy for 6 different excitations (shown by arrows in each panel): $\omega_0=$2.33, 2.38, 2.57, 2.60, 2.65 and 2.77\,eV.
		The light and dark colored lines denotes $I_\text{RL}(\omega,T)$ and $I_\text{RR}(\omega,T)$, respectively.
		The smooth background $f(\omega,T)$ is plotted by black lines.
	}
\end{figure}

Such a high degree of circular polarization for PL cannot 
originate from the bulk bands, which are spin degenerate~\cite{Liu2015PRB}.  
However, all the three surface bands in Fig.~\ref{Fig1}(a) and (b)
exhibit spin-momentum locking,
and
could lead to optical orientation of spins with circularly polarized light.
To identify the electron bands responsible for polarized PL, we study 
the excitation dependence of the peak intensity. 
Figure~\ref{Fig_ExDep} depicts the intensity of polarized PL measured with a right-circular excitation 
with six different energies, as denoted by the arrows in each panel.
As one can see from the figure, 
the polarized PL
peak, whose position is marked by the dashed line,
is absent for excitation energies below 2.4\,eV.
This implies that
the electron and hole bands involved in forming the exciton are separated by at least 2.4\,eV.
By comparing the PL spectra in
the top two panels of Fig.~\ref{Fig_ExDep}, we note however that the spectrum for the  excitation energy of 2.38\,eV 
does not exhibit any visible features 
at the exciton energy, i.e. at 2.3\,eV, whereas weak PL for the excitation energy of 2.33\,eV is 
enhanced at 2.3\,eV. 
Also, the enhancement occurs primarily in the polarized (RR) channel, whereas PL at the excitation energy of 2.38\,eV is not polarized.
We argue that this re-entrant behavior is an indication of
the resonant excitation of dipole-allowed exciton states~\cite{Vinattieri1994PRB}.

Besides the overall intensity, the difference between $I_\text{RL}(\omega,T)$ and $I_\text{RR}(\omega,T)$, 
depicted in Fig.~\ref{Fig_ExDep} by the light and dark colored lines,
respectively, also changes with the excitation energy.
In Fig.~\ref{Fig1}(e) we show the integrated PL intensity difference, 
$\int_{2.0}^{2.5\,eV}[I_\text{RR}(\omega,T)-I_\text{RL}(\omega,T)]d\omega$, versus the excitation energy $\omega_0$.
The polarized PL peak is observed only for excitation energies between 2.6 and 3.0\,eV. 
Comparing the excitation profile with the known band structure of Bi$_2$Se$_3$ [See \textit{SI Appendix}, Sec.~S2], 
we conclude that the only possible interband transition is from RSS to SS2 bands.

\subsection*{Depolarization}
To analyze polarization-preserving PL quantitatively, we decompose
$I_\text{RR}(\omega,T)$ and $I_\text{RL}(\omega,T)$ into two spectral contributions [Fig.~\ref{Fig_ExDep}]: 
(1) a broad unpolarized emission band, $f(\omega,T)$, and (2) a narrower peak that is almost fully polarized, with intensity defined as $\mathcal{L}_{R}(\omega,T)=I_\text{RR}(\omega,T)-f(\omega,T)$. 
We note that $f(\omega,T)$ and $\mathcal{L}_{R}(\omega,T)$ have distinct lineshapes and therefore are likely to have different origins. 
We will henceforth focus on the polarized PL signal, $\mathcal{L}_{R}(\omega,T)$. 
A small fraction of $\mathcal{L}_{R}(\omega,T)$ is also present in the orthogonal polarization emission, $\mathcal{L}_{L}(\omega,T)=I_\text{RL}(\omega,T)-f(\omega,T)= r(T)\mathcal{L}_{R}(\omega,T)$, where \(r(T)\equiv \frac{I_\text{RL}(\omega,T)-f(\omega,T)}{I_\text{RR}(\omega,T)-f(\omega,T)}\) is the depolarization ratio [see Material and Methods]. 

\begin{figure}[t]
	\includegraphics[width=8.7cm]{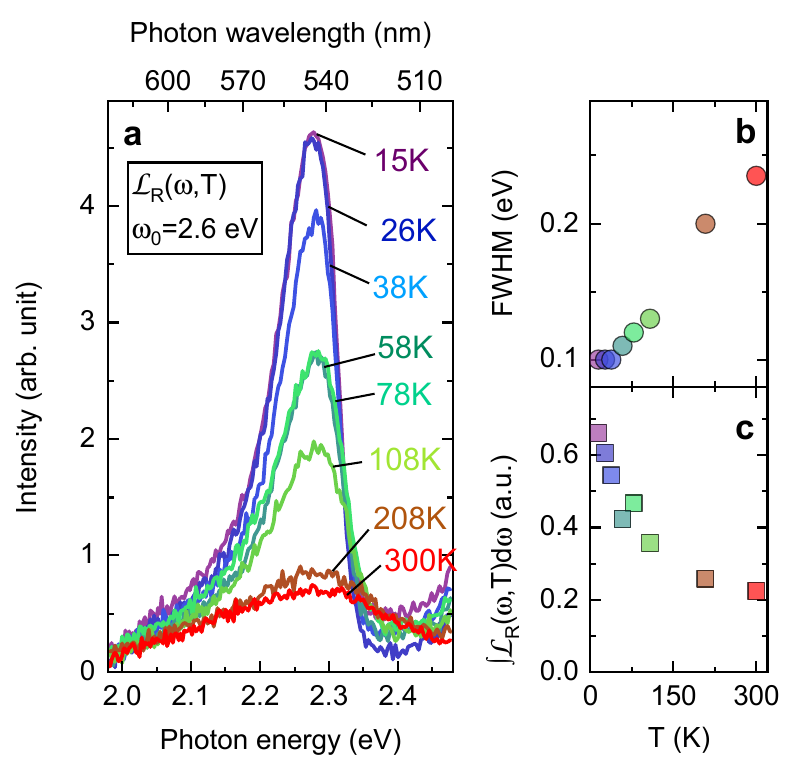}
	\caption{\label{Fig_Tdep}
		(a), The intensity of 
		the polarized PL
		signal, $\mathcal{L}_{R}(\omega,T)$, 
		as a function of the 
		photon energy for various temperatures.
		The temperature dependence of, (b), the full-width-at-half-maximum; and
		(c), the integrated intensity of the polarized PL signal, $\int \mathcal{L}_{R}(\omega,T)\,d\omega$.
		The color coding corresponds to the line colors in (a).
	}
\end{figure}

In Fig.~\ref{Fig_Tdep}, we plot the temperature dependence of $\mathcal{L}_{R}(\omega,T)$,
excited with 2.6\,eV right-circularly polarized light.
While PL is much stronger at 15\,K, emission remains polarized 
even at 300\,K with same $r(T)\approx 0.1$ for all temperatures [\textit{SI Appendix}, Fig.~S6].
This demonstrates that while heating shortens the exciton lifetime, 
it has little impact on polarization of the exciton emission. 

Assuming that the depolarization process 
occurs mostly
during the energy relaxation of the electron and hole 
within the 
corresponding 
bands, we expect $r(T)\rightarrow 0$ as $\omega_0$ approaches the direct surface band gap.
Figure~\ref{Fig1}(f) shows 
that 
$r(T)$
linearly extrapolates
to zero at $\omega_0\approx 2.5$\,eV, which suggests that the band gap should be close to this value. 
This is consistent with a direct transition between the top branches of RSS and SS2, shown by the blue arrow in Fig.~\ref{Fig1}(a).

\subsection*{Surface states}
Now we turn to 
the interpretation of the experimental results.
In general, 
polarization-preserving PL is only possible 
if the spin degeneracy of electron states
is lifted by breaking either time-reversal or inversion symmetries.
Since a bulk Bi$_2$Se$_3$ crystal is non-magnetic and centrosymmetric,  
we argue that observed polarized  PL must be entirely due to surface bands.
In the following, we build a minimal model that explains all the key aspects of the experimental observations,
by considering optically excited electrons and holes in SS2 and RSS bands, respectively.
Based on the first-principle calculation of the electronic band 
structure [Fig.~\ref{Fig1}(b) and \textit{SI Appendix} Fig.~S4], we assert that
both SS2 and RSS states correspond to $J_z=\pm 1/2$ projections 
of the total angular momentum on the $z$-axis and thus can be 
described by $2\times 2$ Pauli matrices.
Also, the mass term in SS2 is by more than a factor of four smaller than the corresponding term in RSS and thus can
be neglected [\textit{SI Appendix}, Sec.~S2].
With these assumptions, we employ the Hamiltonians
\bea
H_{SS2}(\vec{p})&=& \Delta\mathbb{1}_{\sigma_{\text{e}}} + v
(\boldsymbol{\sigma}_{\text{e}}\times\vec{p})\cdot\hat z
,\nonumber\\
H_{RSS}(\vec{p})&=& -\frac{\vec{p}^2}{2m_h}\mathbb{1}_{\sigma_{\text{h}}} -\alpha
(\boldsymbol{\sigma}_{\text{h}}\times\vec{p})\cdot\hat z
\label{Ham_parts}
\eea
to describe the massless Dirac electrons near the SS2 touching point \cite{Fu2009} 
and the massive Rashba holes near the RSS touching point \cite{winkler:book}, respectively.
Here, $\Delta$ is the energy difference between the Dirac points of RSS and SS2,
$v$ is the Dirac velocity, $m_h>0$ is the effective hole mass, $\alpha$ is the Rashba coefficient, 
$\hat z$ is a unit vector normal to the surface,
$\boldsymbol{\sigma}_{\text{e}}$ and $\boldsymbol{\sigma}_{\text{h}}$ are the vectors of Pauli matrices in the SS2 and RSS spin subspaces, respectively, and 
$\mathbb{1}_{\sigma_{\text{e}}}$ and $\mathbb{1}_{\sigma_{\text{h}}}$ are the identity matrices in the same subspaces.
The linear-in-${\bf p}$ terms describe the effect of SOC which locks electron spin at $90^\circ$ to its momentum.
We note that although the RSS band is not topologically protected, in 
contrast to the SS1 and SS2 bands, its band parameters are still 
expected to be universal given atomically smooth and freshly cleaved surfaces, which are realized in Bi$_2$Se$_3$.  

An interacting electron-hole pair is described by a $4\times 4$ two-body Hamiltonian 
\bea 
H_{eh}(\vec{p},\vec{k})&=H_{SS2}\left(\vec{p}+\frac{ 
	\vec{k}}{2}\right) \otimes\mathbb{1}_{\sigma_{\text{h}}}-\mathbb{1}_{\sigma_{\text{e}}}\otimes H_{RSS}\left(-\vec{p}+\frac{\vec{k}}{2}\right)\notag\\
&+\mathbb{1}_{\sigma_{\text{e}}}\otimes\mathbb{1}_{\sigma_{\text{h}}} V(\vec{r}), \label{Ham} 
\eea
where 
$\vec{r}=\vec{r_e}-\vec{r_h}$ is the relative position of the electron and hole, $V(\vec{r})$ describes the Coulomb interaction, $\vec{p}=-i\vec\nabla_{{\bf r}}$ and $\vec{k}$ is the momentum conjugate to $(1/2)\left(\vec{r_e}+\vec{r_h}\right)$.
For $\vec{k}=0$,
the eigenvalues of $H_{eh}(\vec{p},\vec{k})$ have a W-shaped dispersion resembling a multi-layer Mexican hat [cf. Figs.~\ref{Fig1}(d) and S2 in \textit{SI Appendix}].
If both electron and hole bands were massless, a bound state would not be possible.  
However, the two-body bands originating from $H_{eh}(\vec{p},0)$ are bounded from below for any 
values of $v$ and $\alpha$ by the $\vec{p}^2$ term in the RSS band.
Therefore, the Coulomb attraction between electrons and holes lead to excitonic bound states.

\subsection*{Eigenstates and optical transitions}
In what follows, we focus on the case of zero total momentum ($\vec{k}=0$) appropriate for 
direct
optical transitions studied
in this Report.
If $V(\vec{r})$ is axially symmetric,  the $z$ component of the angular momentum of an electron-hole pair
\bea
\hat J_z=\mathbb{1}_{\sigma_{\text{e}}}\otimes\mathbb{1}_{\sigma_{\text{h}}}\left(-i\partial_\phi\right) +\frac 12 \mathbb{1}_{\sigma_{\text{e}}}\otimes\sigma_{\text{h}}^z+\frac 12 \sigma_{\text{e}}^z \otimes\mathbb{1}_{\sigma_{\text{h}}}
\eea
is a good quantum number 
although neither the orbital angular momentum nor spin are 
good quantum numbers 
on their own.  
(Here, $\phi$ is the azimuthal angle of $\vec{p}$.)
Therefore, the eigenstates of Eq.~(\ref{Ham}) can be 
classified by $J_z$.
The Schr\"odinger equation defined by the Hamiltonian in Eq.~(\ref{Ham}) 
can be solved by the following Ansatz for the 4-component spinor wavefunction in the momentum-space representation [\textit{SI Appendix}, Sec.~S1B]:
\bea \label{eq:Ansatz}
\psi(\vec{p})=e^{iJ_z\phi} \left(\psi_1(p) e^{-i\phi},
\psi_2(p),
\psi_3(p),\psi_4(p) e^{i\phi}\right)^T,
\eea
where $p\equiv |\vec{p}|$.
To understand the general properties of the resulting discrete states 
and, in particular, their spin structure, it is instructive to 
replace the interaction potential by a model short-range attraction 
$V(\vec{r})=-\lambda\delta(\vec{r})$, which 
provides a 
reasonable approximation for Coulomb interaction screened by free 
carriers.    
In this case, algebraic equations for amplitudes $\psi_1(p),\dots 
,\psi_4(p)$ have non-trivial solutions for an infinitesimally small 
$\lambda$, but only for states with
$J_z=0$ and $J_z=\pm 1$ [\textit{SI Appendix, Sec.~S1A}].  
The bound states with $J_z=\pm 1$, labeled as $\ket{J_z=\pm 1}$ in Fig.~\ref{Fig1}(d), are doubly degenerate, whereas the two states with $J_z=0$, labeled as $\ket{J_z=0}$, have different energies\footnote{Strictly speaking,  exciton states have to be classified within the 
surface symmetry group, which is $C_{6v}$ 
for actual Bi$_2$Se$_3$ \cite{Li2013Mar} or $C_{\infty v}$ for a rotationally-invariant Hamiltonian in Eq.~(\ref{Ham}).
Inspecting the exciton wave functions in Eq.~(\ref {eq:Ansatz}), we find that the two $\mathinner {|{J_z=0}\delimiter "526930B }$ states are fully symmetric with respect to all symmetry operators of 
both $C_{6v}$ and $C_{\infty v}$groups, and therefore 
belong
to the $A_1$ irreducible representation. On the other hand, the doubly degenerate $\mathinner {|{J_z=\pm 1}\delimiter "526930B }$ states belong to the $E_1$ representation, which transform as an in-plane electric dipole.}.
Within the backscattering geometry of our experiment, circularly 
polarized light can only 
produce excitations with $\Delta J_z=\pm 1$.
Assuming no cross-relaxation between $\ket{J_z=\pm 1}$ and 
$\ket{J_z=0}$ states, we expect 
a single 
PL peak 
arising from recombination of
the $\ket{J_z=\pm 1}$ exciton, which is 
consistent with the data [Fig.~\ref{Fig1}(c)].

The above argument is suitable for explaining 
polarized PL 
excited by photons with energies
close to the Mexican-hat minimum of Fig.~\ref{Fig1}(d).
One could expect 
that 
scattering by phonons
couples the $\ket{J_z=+1}$ and 
$\ket{J_z=-1}$ states for energies above the minimum, which would 
cause an increase of $r(T)$ with $\omega_0$. 
However, we see only a moderate increase of $r(T)$ even if 
$\omega_0$ is about 300\,meV above the Mexican-hat minimum [Fig.~\ref{Fig_ExDep}].
To explain the preservation of 
optical orientation during energy relaxation,
we note that a transition between the $\ket{J_z=\pm 1}$ states do not conserve the $z$ component of the angular momentum, 
and thus requires scattering by non-symmetric bosons or by a magnetic impurity.
It is known that non-symmetric surface phonons in Bi$_2$Se$_3$ are weak~\cite{Kung2017,Sobota2014}, leaving $J_z$ approximately conserved during the energy relaxation.
It would be interesting to study in the future the interaction between the chiral exciton and other more exotic collective modes, such as the Dirac plasmons and chiral spin modes~\cite{Raghu2010,Politano2015,Kung2017PRL}.
Importantly, the $\ket{J_z=\pm 1}$ exciton states can also be resonantly populated with circularly polarized 2.3\,eV excitation [Fig.~\ref{Fig_ExDep}],
which suggests that the exciton states are dipole-allowed and thus confirms the proposed model.

\subsection*{Bound state energies}
\begin{figure}[t]
	\includegraphics[width=8.7cm]{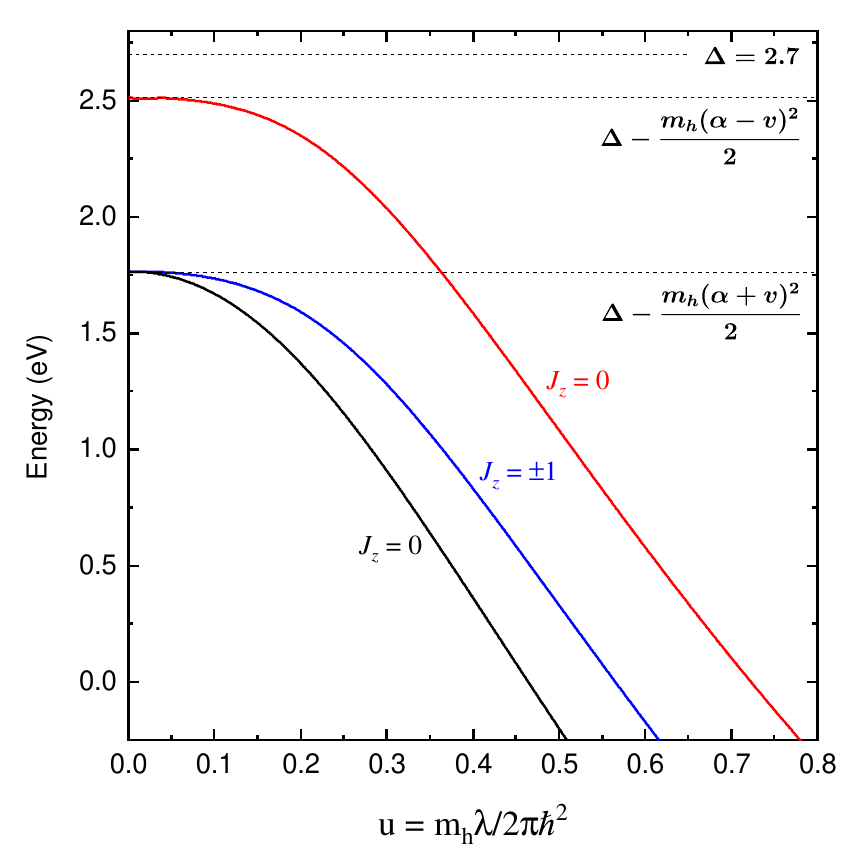}
	\caption{\label{fig:boundstates}
	\textbf{Bound state energies of excitons with $J_z=\pm 1$ (blue) and $J_z=0$ (red and black)}
		obtained by numerically diagonalizing Eq.~(\ref{Ham}) with $V(\vec{r})=-\lambda\delta(\vec{r})$. 
		The band structure parameters are taken from fitting the ARPES data of Refs.~\cite{Sobota2013} and \cite{Soifer2017} into the spectrum of Eq.~(\ref{Ham_parts}): $\Delta=2.7$\,eV, $v=2.0$\,eV\AA, $\alpha=5.2$\,eV\AA, $m_h=0.036$\,eV$^{-1}$\AA$^{-2}$ [\textit{SI Appendix}, Sec.~S2], and $u=m_h\lambda/2\pi\hbar^2$.
	} 
\end{figure}

The theoretical model described above allows one to extract 
quantitative characteristics of the exciton spectra. 
The exciton energies for  a short-range interaction as functions of the dimensionless coupling constant $u=m_h\lambda/2\pi\hbar^2
$ are 
shown in Fig.~\ref{fig:boundstates}.
With the band parameters extracted from ARPES data 
\cite{Sobota2013,Soifer2017}, 
one finds for the absorption edge in the $J_z=\pm 1$ channel $E^{\pm 1}_g=\Delta-m_h(\alpha+v)^2/2\approx 1.8$\,eV, which is somewhat smaller but comparable to the observed value of $2.48$\,eV. 
For a more realistic case of the Coulomb interaction (which is assumed to be weak),
the binding energy 
can be estimated as 
$\epsilon_{\pm 1}-E_g^{\pm 1}=-4\text{Ry}^*\ln^2\left[2 m_h(\alpha+v)a_{B}/
\text{e}^{2}\hbar\right]$~\cite{chaplik:2006,grimaldi:2008},
where $\text{Ry}^*=m_he^4/2\hbar^2\varepsilon^2_{\text{eff}}\approx 0.02$\,eV is the effective Rydberg, $\varepsilon_{\text{eff}}\approx 13$ is  the effective dielectric constant 
of semi-infinite
Bi$_2$Se$_3$
for frequencies
above the
topmost phonon mode, $a_B=\hbar^2\varepsilon_{\text{eff}}/m_he^2$ is the effective Bohr radius, and 
$\text{e}=2.718\dots$ is the base of the natural logarithm.  
The (large) logarithmic factor in the bound state energy 
arises because massive holes 
with energies close to the minimum of the Rashba spectrum exhibit an effectively one-dimensional (1D) motion \cite{chaplik:2006}.
In 1D,  the bound state energy in a weak potential $U(x)$ is proportional to $\left[\int dx U(x)\right]^2$ \cite{Landau:77}, 
hence the $\ln^2$ factor for the $1/x$ potential.
A more accurate result can be obtained by numerical solution of the Schr\"odinger equation \cite{grimaldi:2008} which gives $0.22$\,eV for the bound state energy, whereas the observed value is $0.2$\,eV. 
We thus conclude that our theoretical model is in quantitative agreement with the data.

We note that while $\alpha$ and $v$ may vary slightly
from sample to sample, the chiral exciton energy depends only logarithmically on the band structure parameters,
at least as long as the Coulomb attraction between electron and hole is sufficiently weak.
Furthermore, it is known from ARPES measurements, nonlinear optics, and first principle calculations that, while the 
the position of the
Fermi level is very sensitive to surface preparation, the 
surface states are rather robust against nonmagnetic dopants~\cite{Chen2009,Hsieh2011A,Koleini2013PRL}. 
In our case, the surface states composing the chiral exciton are 
far away from the Fermi level, and thus should be even less sensitive to surface contamination.
This naturally explains the reproducibility of the observed features between samples.

\subsection*{Conclusions}
We used polarization-resolved photoluminescence (PL) spectroscopy to study the secondary emission from the surface states of an archetypical topological insulator Bi$_2$Se$_3$. 
When the crystal is excited with 2.5--2.8\,eV circularly polarized light, we detect emission of the same polarization at 2.3\,eV. 
Polarization of emitted light is preserved even if the excitation energy is hundreds of meVs above the emission threshold energy. 
We assign such emission as resulting from recombination of exciton states:
 \textit{chiral excitons}. 
We propose that chiral excitons are made of  (topologically protected) massless electrons and massive holes, both residing on the surface of  Bi$_2$Se$_3$ and characterized by chiral spin textures. 
The exciton states can be characterized by the eigenvalues of the out-of-plane total angular momentum, $J_z$. 
Based on the results of our theoretical model, we identify the doublet of degenerate states with $J_z=\pm 1$ as being responsible for observed polarization-preserving PL. 
The most surprising finding is that polarization of chiral exciton PL is preserved up to room temperature and robust with respect to chemical substitution, which we attribute to the weakness of spin-flip scattering between surface states with opposite helicity. 
In this way, chiral excitons are fundamentally different from other known excitons that also preserve helicity~\cite{Xu2014,Planel1984book}. 
Controlled optical orientation of chiral surface excitons may facilitate new photonics and optoelectronics applications of topological insulators.

\begin{acknowledgments}
    We are grateful for technical support from B.S.~Dennis and A.~Rustagi, and discussions with L.S.~Levitov, Z.~Lu, E.I.~Rashba, Z.X.~Shen, D.~Smirnov, J.A.~Sobota, H.~Soifer and D.~Vanderbilt.
    We acknowledge support from NSF DMR-1104884 (G.B., H.-H.K and A.L.), 
    NSF DMR-1720816 (D.L.M.), and NSF Grant No. DMR-1629059 (to S.-W.C. and X.W.).
\end{acknowledgments}


\section*{Material \& Methods}
\subsection*{Material growth}
All data presented in the main text are collected from bulk single crystals grown by modified Bridgman method.
Mixtures of high-purity bismuth (99.999\%) and selenium (99.999\%) with the mole ratio $\text{Bi}:\text{Se}=2:3$ were heated up to 870\,$^\circ$C in sealed vacuum quartz tubes for 10 hours, and then slowly cooled to 200\,$^\circ$C with rate 3\,$^\circ$C/h, followed by furnace cooling to room temperature.

\subsection*{Experimental setup}
The crystals were cleaved prior to cool down in a glove bag filled with nitrogen gas, and were transferred into a continuous flow liquid helium optical cryostat without exposure to atmosphere.
A solid state laser was used for 2.33\,eV (532\,nm) excitation, a diode laser was used for 2.77\,eV (447\,nm) excitation, and a Kr$^{+}$ ion laser was used for all other excitations, with laser spot size roughly $50\times 50\,\mu m^2$.
The power density on the sample is kept below 0.7\,kW/cm$^2$, and all temperatures shown were corrected for laser heating with 1\,K/mW.
The polarized secondary emission was analyzed and collected by a custom triple-grating spectrometer with a liquid nitrogen cooled CCD detector. 

The intensity $I_{\mu\nu}(\omega,T)$, was corrected for the laser power and spectral response of the spectrometer and CCD, where $\mu$ ($\nu$) denotes the direction of incident (collected) photon polarization, $\omega$ is energy and $T$ is temperature.
The scattering geometries used in this experiment are denoted as $\mu\nu=$RR, RL, XX and YX.
$\text{R}=\text{X}+i\text{Y}$ and $\text{L}=\text{X}-i\text{Y}$ denotes the right- and left-circular polarizations, respectively, where X (Y) denotes linear polarization parallel (orthogonal) to the plane of incidence.
Here, we follow the ``spectroscopy convention'' for the ``handedness'' of circularly polarized light. That is, the right and left polarization refers to the angular momentum measured in the lab frame, rather than the helicity of photon. 

\subsection*{Photoluminescence background subtraction}
With right circularly polarized excitation, the measured PL intensities can be decomposed into two parts:
\begin{align}
	I_\text{RR}(\omega,T)&=\mathcal{L}_{R}(\omega,T) + f(\omega,T) \nonumber\\
	I_\text{RL}(\omega,T)&=\mathcal{L}_{L}(\omega,T) + f(\omega,T)
\end{align}

Where $\mathcal{L}_{R}(\omega,T)$ and $\mathcal{L}_{L}(\omega,T)$ denotes right and left circularly polarized PL, respectively, and $f(\omega,T)$ denotes the featureless unpolarized broad background.
We assume an energy independent depolarization ratio $r(T)=\frac{\mathcal{L}_{L}(\omega,T)}{\mathcal{L}_{R}(\omega,T)}$.
Inserting $r(T)$ into the above expression of $I_\text{RL}(\omega,T)$, we can write the unpolarized emission as:
\begin{equation}
f(\omega,T)=\frac{I_\text{RL}(\omega,T)-r(T)\cdot I_\text{RR}(\omega,T)}{1-r(T)} .
\end{equation}
Then, $r(T)$ is determined by minimizing sharp features in $f(\omega,T)$ around the 2.3\,eV PL peak.
The circularly polarized PL can be calculated knowing $r(T)$,
\begin{equation}
\mathcal{L}_{R}(\omega,T)=\frac{I_\text{RR}(\omega,T)-I_\text{RL}(\omega,T)}{1-r(T)} .
\end{equation} 

%
%

\clearpage
\onecolumngrid
\appendix
\setstretch{1.5}
\renewcommand{\thefigure}{S\arabic{figure}}
\addtocounter{figure}{-5}
\renewcommand{\theequation}{S\arabic{equation}}
\addtocounter{equation}{-7}
\renewcommand{\thetable}{S\arabic{table}}
\begin{center}
    \textbf{\Large
        Supplementary Information for:\\
        Observation of Chiral Surface Excitons in a Topological Insulator Bi$_2$Se$_3$} \par
\end{center}
\section{1.~Theory of two-dimensional chiral excitons}\label{SM:theory}
\subsection{A.~Scattering-matrix approach} \label{sec:scat_matrix}

In this section, we analyze the 
chiral excitonic states in the $T$-matrix formalism, in which the bound states of Hamiltonian Eq.~(2) in the Main Text (MT) are found by solving
\begin{align}
1&=-\lambda Z(\epsilon,\vec{k}),\notag\\
Z(\epsilon,\vec{k})&=\sum_{\vec{p}}\frac1{\epsilon-\Delta-\frac{\vec{p}_-^2}{2m_{\text{h}}}-v(\boldsymbol{\sigma}_{\text{e}}\times\vec{p}_+)\cdot\hat{\vec{z}}+\alpha(\boldsymbol{\sigma}_{\text{h}}\times\vec{p}_-)\cdot\hat{\vec{z}}+i0},\label{Eq:Tmatrix}
\end{align}
where $\vec{p}_{\pm}=\vec{p}\pm \vec{k}/2$ and $\sum_\vec{p}$ is a shorthand for $\int d^2p/(2\pi\hbar)^2$.
As in MT, we focus on the $\vec{k}=0$ case.  It is instructive to rewrite the expression under $\sum_{\vec p}$  in Eq.~(\ref{Eq:Tmatrix}) using Schwinger proper time parameterization
\be\label{eq:proper_time_1}
\frac1{\epsilon-\Delta-\frac{p^2}{2m_{\text{h}}}-v(\boldsymbol{\sigma}_{\text{e}}\times\vec p)\cdot\hat{\vec{z}}+\alpha(\boldsymbol{\sigma}_{\text{h}}\times\vec p)\cdot\hat{\vec{z}}+i0}
=-\int_0^\infty dt\, e^{-t(\Delta+p^2/2m_h-\epsilon +v(\boldsymbol{\sigma}_{\text{e}}\times\vec p)\cdot\hat{\vec{z}}-\alpha(\boldsymbol{\sigma}_{\text{h}}\times\vec p)\cdot\hat{\vec{z}})}.
\ee
The expression above is particularly well-suited for performing angular averaging over $\vec p$. 
We first simplify the exponent by factoring it into parts depending on $\boldsymbol{\sigma}{_{\text{e}}}$ and $\boldsymbol{\sigma}_{\text{h}}$. Also, we assume that $\alpha>0$ and $v>0$ throughout the Supporting Information (SI).
Next we use the identities:
\begin{align}
e^{-t v(\boldsymbol{\sigma}_{\text{e}}\times\vec p)\cdot\hat{\vec{z}}}&=\cosh(t vp)
-\sinh(t vp)(\boldsymbol{\sigma}_{\text{e}}\times\hat{\vec p})\cdot\hat{\vec{z}}, \nn
e^{+t \alpha(\boldsymbol{\sigma}_{\text{h}}\times\vec p)\cdot\hat{\vec{z}}}&=\cosh(t \alpha p)
+\sinh(t \alpha p)(\boldsymbol{\sigma}_{\text{h}}\times\hat{\vec p})\cdot\hat{\vec{z}},
\end{align}
where $\hat{\vec p}=\vec p/p$ is the unit vector along $\vec{p}$.
Crucially, the terms $(\boldsymbol{\sigma}_{\text{e}}\times\hat{\vec p})\cdot\hat{\vec{z}}$ and $(\boldsymbol{\sigma}_{\text{h}}\times\hat{\vec p})\cdot\hat{\vec{z}}$ are odd in $\vec p$ and vanish upon averaging over  the  orientations of $\vec p$. 
Keeping only the terms which are even in $\vec p$, we are left with
\be\label{eq:proper_time_2}
-\int_0^\infty dt\, e^{-t(\Delta+p^2/2m_h-\epsilon)}
\lb \cosh(t vp) \cosh(t \alpha p)
- \sinh(t vp) \sinh(t \alpha p)((\boldsymbol{\sigma}_{\text{e}}\times\hat{\vec p})\cdot\hat{\vec{z}})\otimes((\boldsymbol{\sigma}_{\text{h}}\times\hat{\vec p})\cdot\hat{\vec{z}}) \rb.
\ee
Upon averaging over angular part of $\vec p$, the term $((\boldsymbol{\sigma}_{\text{e}}\times\hat{\vec p})\cdot\hat{\vec{z}})\otimes ((\boldsymbol{\sigma}_{\text{h}}\times\hat{\vec p})\cdot\hat{\vec{z}})$ yields 
\be
\Xi=\sigma_{\text{e}}^+\otimes\sigma_{\text{h}}^-+\sigma_{\text{e}}^-\otimes\sigma_{\text{h}}^+,\label{Xi}
\ee
where $\sigma_{\text{e}}^\pm=(\sigma_{\text{e}}^x\pm i\sigma_{\text{e}}^y)/2$ and similarly for $\sigma_{\text{h}}^\pm$. 
The exciton spin wave functions are described by the eigenvectors of the operator $\Xi$, which has a double degenerate zero eigenvalue and nondegenerate eigenvalues $\pm 1$.
Hence, $Z(\epsilon,\vec k=0)$ has the matrix form $A\mathbb{1}_{\sigma_{\text{e}}}\otimes\mathbb{1}_{\sigma_{\text{h}}}+B\Xi$, where $A$ and $B$ are scalars.
In the eigenbasis of $\Xi$, it is diagonal, with entries $A$ and $A$ for the $\xi=0$ case, and entries $A+B$ and $A-B$ for the $\xi=1$ and $\xi=-1$ cases, respectively.

For $\xi=0$, the term $\sinh(t vp) \sinh(t \alpha p)$ in Eq.~(\ref{eq:proper_time_2}) is absent. Evaluating the integral over $t$, we obtain the angle-averaged quantity
\be\label{eq: all 4}
\la \frac1{\epsilon-\Delta-\frac{ p^2}{2m_{\text{h}}}-v(\boldsymbol{\sigma}_{\text{e}}\times\vec p)\cdot\hat{\vec{z}}+\alpha(\boldsymbol{\sigma}_{\text{h}}\times\vec p)\cdot\hat{\vec{z}}+i0} \ra =
\frac14\sum_{\eta_e,\eta_h=\pm 1}\frac1{\epsilon-\Delta-\frac{ p^2}{2m_{\text{h}}}-\eta_e vp+\eta_h \alpha p+i0}.
\ee
For $\xi=+1$ we replace the term $((\boldsymbol{\sigma}_{\text{e}}\times\hat{\vec p})\cdot\hat{\vec{z}})\otimes ((\boldsymbol{\sigma}_{\text{h}}\times\hat{\vec p})\cdot\hat{\vec{z}})$ by unity and  find
\be\label{eq: ++,--}
\la\frac1{\epsilon-\Delta-\frac{ p^2}{2m_{\text{h}}}-v(\boldsymbol{\sigma}_{\text{e}}\times\vec p)\cdot\hat{\vec{z}}+\alpha(\boldsymbol{\sigma}_{\text{h}}\times\vec p)\cdot\hat{\vec{z}}+i0} \ra 
=\frac12\sum_{\eta=\pm 1}\frac1{\epsilon-\Delta-\frac{ p^2}{2m_{\text{h}}}-\eta (v-\alpha)p+i0}.
\ee
Similarly, for $\xi=-1$ we obtain
\be\label{eq: -+,+-}
\la \frac1{\epsilon-\Delta-\frac{ p^2}{2m_{\text{h}}}-v(\boldsymbol{\sigma}_{\text{e}}\times\vec p)\cdot\hat{\vec{z}}+\alpha(\boldsymbol{\sigma}_{\text{h}}\times\vec p)\cdot\hat{\vec{z}}+i0} \ra =\frac12\sum_{\eta=\pm 1}\frac1{\epsilon-\Delta-\frac{ p^2}{2m_{\text{h}}}-\eta (v+\alpha)p+i0}.
\ee

The treatment above is based on solving an effective two-body problem. Alternatively, one can start with a many-body formulation, in which
electrons in the conduction (SS2) and valence (RSS) bands are treated on the same footing, except for the dispersions in the valence bands are of the opposite sign to that in the conduction ones. The original interaction is this formulation is a {\em repulsive} density-density interaction between electrons. The Hamiltonian of the many-body problem is written as 
\bea
H&=&\sum_{ \vec{p},\sigma_{\text{e}},\sigma_{\text{h}}}a^{\dagger}_{\vec p\sigma_{\text{e}}} a_{\vec p\sigma_{\text{e}}}^{\phantom{\dagger}} H_{SS2}(\vec{p}) \otimes\mathbb{1}_{\sigma_{\text{h}}}+ \sum_{ \vec{p},\sigma_{\text{e}},\sigma_{\text{h}}}b^{\dagger}_{\vec{p}\sigma_{\text{h}}} b_{\vec{p}\sigma_{\text{h}}}^{\phantom{\dagger}}  \mathbb{1}_{\sigma_{\text{e}}}\otimes H_{RSS}(\vec{p}) \nn
&+&\sum_{\vec{p},\vec{p}',\vec{k},\sigma_{\text{e}},\sigma_{\text{h}}} V(\vec{k}) \mathbb{1}_{\sigma_{\text{e}}}\otimes\mathbb{1}_{\sigma_{\text{h}}}
a^{\dagger}_{\vec{p}+\vec{k}\sigma_{\text{e}}} a_{\vec{p}\sigma_{\text{e}}}^{\phantom{\dagger}}b^{\dagger}_{\vec{p}'-\vec{k}\sigma_{\text{h}}} b_{\vec{p}'\sigma_{\text{h}}}^{\phantom{\dagger}},\label{ham}
\eea
where $H_{SS2}(\vec{p})$ and $H_{RSS}(\vec{p})$ are the single-particle Hamiltonians given in Eq.~(1) of the MT and $V(\vec{k})>0$ is the {\em repulsive} interaction potential.

\begin{figure}[t]
    \centering
    \includegraphics[width=16cm]{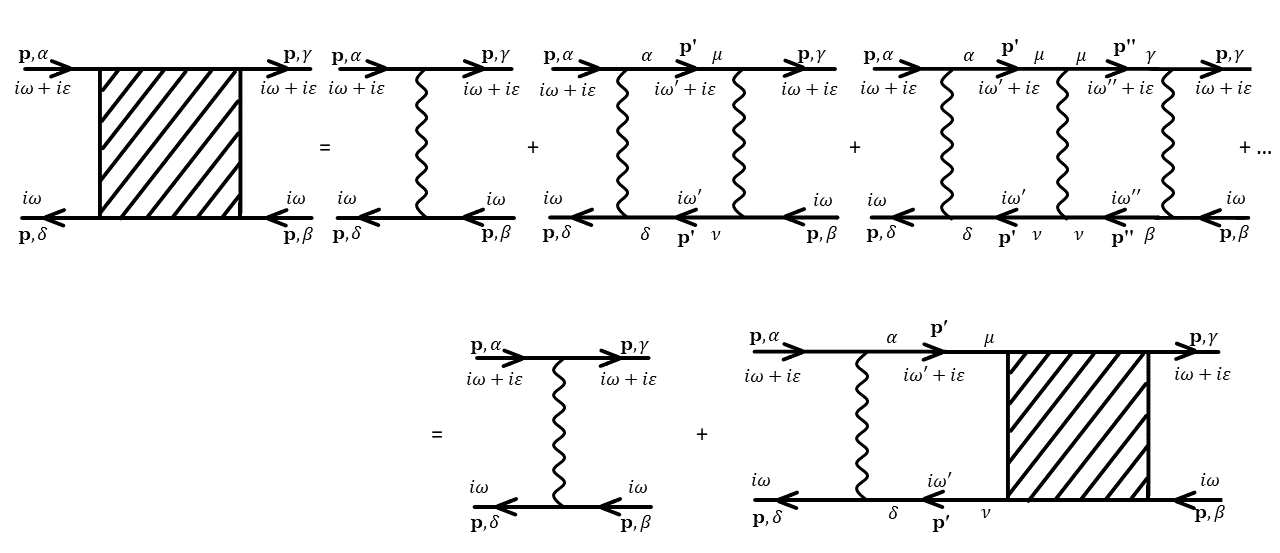}
    \caption{\label{Ladder_Diagram}
        A graphical representation of the integral equation for the scattering vertex $\Gamma_{\alpha\beta,\gamma\delta}$.}
\end{figure}

The interaction vertex corresponding to the Hamiltonian (\ref{ham}) satisfies the integral equation represented graphically in Fig.~(\ref{Ladder_Diagram}).
Because electron states are always empty while hole states are always occupied, the problem is equivalent to that in vacuum and there are no other diagrams except for ladder ones. For a point-like interaction $V(\vec{k})=\lambda>0$, the integral equation is reduced to an algebraic one
\be\label{eq:Gamma}
\Gamma_{\alpha\beta,\gamma\delta}=\lambda
\delta_{\alpha\gamma}
\delta_{\beta\delta}
-\lambda\,
\sum
_{\mu,\nu}
\int
\frac{d^2 p'}{(2\pi\hbar)^2} T\sum_{\omega_m}\, G^e_{\alpha\mu}(\vec p',i\omega_m+i\varepsilon_n)G^h_{\nu\delta}(\vec p',i\omega_m)\,\Gamma_{\mu\beta,\gamma\nu},
\ee
Here,
${G}^{e,h}(\vec{p},i\omega)$ are the (Matsubara) Greens functions for electrons and holes,  respectively, given by
\bea\label{eq: G}
{G}^{e}(\vec p,i\omega_m)&=&\sum_{\eta_e=+,-}\frac{1}{2}\lb\mathbb{1}_{\sigma_{\text{e}}}+\eta_e
(\boldsymbol{\sigma}_{\text{e}} \times\hat{\vec p})\cdot\hat{\vec z}\rb g^e_{\eta_e}(\vec p,i\omega_m),\nn
{G}^{h}(\vec p,i\omega_m)&=&\sum_{\eta_h=+,-}\frac{1}{2}\lb\mathbb{1}_{\sigma_{\text{h}}}+\eta_h
(\boldsymbol{\sigma}_{\text{h}} \times\hat{\vec p})\cdot\hat{\vec z}\rb g^h_{\eta_{\text{h}}}(\vec p,i\omega_m),
\eea
where 
\be\label{eq: g}
g^e_{\eta_e}(\vec p,i\omega_m) = \frac{1}{i\omega_m-\epsilon^e_{\eta_e}(\vec p)+\mu},\qquad g^h_{\eta_{\text{h}}}(\vec p,\omega) = \frac{1}{i\omega_m-\epsilon^h_{\eta_{\text{h}}}(\vec p)+\mu},
\ee
$\mu>0$ is the chemical potential and $\epsilon^e_{\eta_e}$, $\epsilon^h_{\eta_{\text{h}}}$ are the eigenvalues of the corresponding single-particle Hamiltonians:
\be\label{eq: epsilon}
\epsilon^e_{\eta_e}(\vec p)=\Delta+\eta_e |v|p,\qquad\epsilon^h_{\eta_{\text{h}}}(\vec p)=-\lb\frac{p^2}{2m_{\text{h}}}+\eta_h|\alpha| p\rb
\ee
with $\eta_e,\eta_h=\pm 1$. 

Equation (\ref{eq:Gamma}) is solved by expanding  $\Gamma_{\alpha\beta,\gamma\delta}$ over the tensor products of spin spaces
\be\label{eq: decomposition}
\Gamma_{\alpha\beta,\gamma\delta}=\sum_{i,j=0}^3 C_{ij} (\sigma_{\text{e}}^{i})_{\alpha\gamma}(\sigma_{\text{h}}^{j})_{\beta\delta},
\ee
where it is understood that ${\sigma_{\text{e}}}^0
=\mathbb{1}_{\sigma_{\text{e}}}$ and 
${\sigma_{\text{h}}}^0=\mathbb{1}_{\sigma_{\text{h}}}$. 
Substituting Eq.~(\ref{eq: decomposition}) into Eq.~(\ref{eq:Gamma}) and projecting out the Pauli matrices, we find for the matrix form of $\Gamma$
\be
\Gamma
=\frac{D_1}{N(4M^2-N^2)}\mathbb{1}_{\sigma_{\text{e}}}\otimes\mathbb{1}_{\sigma_{\text{h}}}
+\frac{D_2}{(4M^2-N^2)}
\,\Xi+ \frac{D_3}{N(4M^2-N^2)}\sigma_{\text{e}}^3\otimes\sigma_{\text{h}}^3,
\label{Gamma_result}
\ee
Here, matrix $\Xi$ is given in Eq.~(\ref{Xi}), $D_1=(8M^2-4N^2)\lambda$, $D_2=-4M\lambda$, $D_3=-8M^2\lambda$,
\begin{eqnarray}
M=-\frac{\lambda}{2}
\sum_{\eta_e,\eta_h=\pm1}\eta_e\eta_h\Pi_{\eta_e,\eta_{\text{h}}},\qquad
N=4+\lambda\sum_{\eta_e,\eta_h=\pm 1}\Pi_{\eta_e\eta_{\text{h}}}\label{evalues},
\end{eqnarray}
and  
\be\label{eq: bubble0}
\Pi_{\eta_e\eta_{\text{h}}}=
\int \frac{d^2 p'}{(2\pi\hbar)^2} T\sum_{\omega_m}g^e_{\eta_e}(\vec p',i\omega_m+i\varepsilon_n)g^h_{\eta_{\text{h}}}(\vec p',i\omega)
\ee
is the convolution of the Green's functions forming the ``rungs'' in the ladder series. (Up to an overall sign, $\Pi_{\eta_e\eta_{\text{h}}}$ is the polarization bubble
formed by electrons from bands $\eta_e$ and $\eta_h$.)
Summing over $\omega_m$ in Eq.~(\ref{eq: bubble0}) and performing analytic continuation $i\varepsilon_n\to \epsilon+i0^+$, we obtain at $T=0$
\be\label{eq: bubble}
\Pi_{\eta_e\eta_{\text{h}}}=\int^\Lambda_0 \frac{d p'p'}{2\pi\hbar^2}\frac{1}{\epsilon-\epsilon^e_{\eta_e}(\vec p')+\epsilon^h_{\eta_{\text{h}}}(\vec p')+i0^+},
\ee
where $\Lambda$ is the ultraviolet cutoff imposed to regularize the logarithmically divergent integral.

Poles of $\Gamma$ give the bound states energies. From Eq.~(\ref{Gamma_result}) we see that the poles  of $\Gamma$
are determined by three equations
\begin{eqnarray}
N=0;\qquad
N=\pm 2M\label{explicit_evalues}.
\end{eqnarray}
Thus, Eq.~(\ref{explicit_evalues}) reads 
\begin{subequations}
    \begin{eqnarray}
    1+\frac{\lambda}{4}\left(\Pi_{++}+\Pi_{--}+\Pi_{-+}+\Pi_{+-}\right)=0,\qquad\xi=0,\
    \label{E1}\\
    1+\frac{\lambda}{2}\left(\Pi_{++}+\Pi_{--}\right)=0,\qquad\xi=+1,
    \label{A1}\\
    1+\frac{\lambda}{2}\left(\Pi_{-+}+\Pi_{+-}\right)=0,\qquad\xi=-1.
    \label{A2}
    \end{eqnarray} 
\end{subequations}

The three equations above are equivalent to Eqs.~(\ref{eq: all 4}), (\ref{eq: ++,--}) and (\ref{eq: -+,+-}) obtained within the  $T$-matrix approach [also, cf. Sec.~1B]. Also,  because $\sigma_{\text{e}}^3\otimes\sigma_{\text{h}}^3$ commutes with $\Xi$, the eigenvalues of $\Gamma$ can be parameterized by those of $\Xi$. 
The effective attraction, which is necessary for forming a bound state, 
arises because the combinations of $\Pi_{ij}$ in Eq.~(\ref{E1}-\ref{A2}) are negative and real is certain energy intervals; therefore, equations have solutions for $\lambda>0$. This a familiar effect of nesting: the original repulsive interaction is replaced with an effective attraction between electron and hole states.

In Sec.~1B, it will be shown that the bound state corresponding to $\xi=0$ is a doubly degenerate one with projections of the total angular momentum on the $z$-axis equal to $J_z=\pm 1$, while the bound states corresponding to $\xi=\pm 1$ are two non-degenerate states with $J_z=0$.
Getting somewhat ahead, we will adopt the notations for the bound states based on the values of $J_z$ rather than of $\xi$.

Continuing with the $T$-matrix approach, we write the $T$-matrix as
\begin{eqnarray}
T = \frac{\lambda}{1+\lambda Z(\epsilon,\vec k)}.
\end{eqnarray}
The imaginary parts of the $T$-matrix in the different symmetry channels are given by 
\bea
T''_{J_z}&=&
\frac{\lambda X_{J_z}''}{(1+\lambda X_{J_z}')^{2}+(\lambda X_{J_z}'')^{2}},
\eea
where $J_z=0,\pm 1$ (it is understood that $J_z=0$ corresponds to two eigenstates labeled as $0^\pm$ below)
and
\begin{subequations}
    \bea 
    X_{\pm 1}
    &=&\frac{1}{4}\left(\Pi_{++}+\Pi_{-+}+\Pi_{+-}+\Pi_{--}\right),\label{E}\\
    X_{
        0^+}&=&\frac{1}{2}\left(\Pi_{++}+\Pi_{--}\right),\label{A1}\\
    X_{0^-
    }&=&\frac{1}{2}\left(\Pi_{-+}+\Pi_{+-}\right).\label{A2}
    \eea
\end{subequations}
with $Y'\equiv \Re Y$, $Y''\equiv \Im Y$ for all quantities. Table~\ref{table:pis} gives explicit expressions for  the real and imaginary parts of $\Pi_{\eta_e\eta_{\text{h}}}$.
The interval of energies in which $X''_{J_z}\neq 0$ correspond to the continua of transitions between electron and hole states. The energy below which $X''_{J_z}$ vanishes defines the absorption edge in a given 
channel, $E^{J_z}_g$.  An inspection of Table~\ref{table:pis} shows that  
\begin{subequations}
    \bea
    &&E^{0^-}_g=\Delta-\frac 12 m_h(\alpha-v)^2;\\
    &&E^{\pm 1}_g=E^{0^+}_g=\Delta-\frac 12 m_h(\alpha+v)^2.
    \eea
\end{subequations}
Bound states can be formed at energies below the absorption edges. Therefore, the bound state in a given channel is possible if $E^{J_z}_g>0$ in this channel. According to the numerical values of the bandstructure parameters  (cf. Sec.~II), these conditions is satisfied in all symmetry channels.

\begin{figure}[t]
    \centering
    \includegraphics[width=12cm]{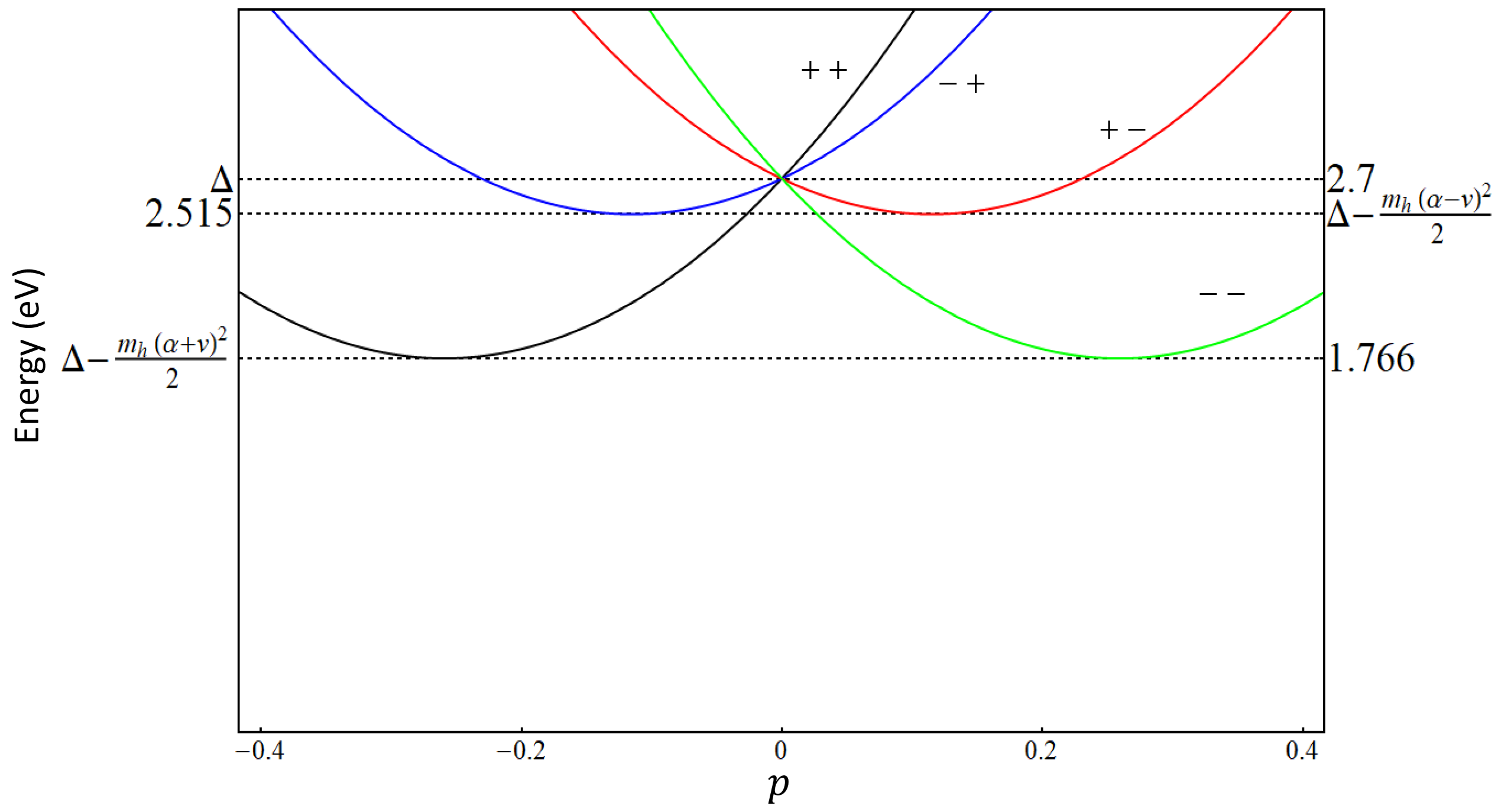}
    \caption{\label{Parabolas}
        Construction of the absorption edges.}
\end{figure}

In fact, the absorption edges can be deduced without a lengthy calculation. Indeed, a finite imaginary part of $\Pi_{\eta_e\eta_{\text{h}}}$ occurs
if the integrand has a pole, i.e., if the following equation is satisfied
\bea\epsilon=\epsilon_{\eta_e}(p)-\epsilon_{\eta_{\text{h}}}(p)=\Delta+\frac{p^2}{2m_{\text{h}}}+(\eta_e v+\eta_h\alpha) p.\eea 
The four parabolas on the right-hand side of this equation are shown in Fig.~\ref{Parabolas}. The $
J_z=0^-$ channel contains combinations $
\left\{\eta_e=1,\eta_h=-1\right\}$ and $\left\{\eta_e=-1,\eta_h=1\right\}$ [cf. Eq.~(\ref{A2})], and the corresponding parabolas have minima of  $\Delta-m_h(\alpha-v)^2/2$. For $\epsilon<\Delta-m_h(\alpha-v)^2/2$, the denominator of the integrand in Eq.~(\ref{eq: bubble}) is negative and there is no pole. Therefore, the energy $\Delta-m_h(\alpha-v)^2/2$ indeed corresponds to the absorption edge in the $
J_z=0^-$ channel. Likewise, the $
J_z=\pm 1$ and $
J_z=0^+$ channels contain combinations $\left\{\eta_e=1,\eta_h=1\right\}$ and $\left\{\eta_e=-1,\eta_h=-1\right\}$. To avoid poles in these channels, one needs to go below the minima of the corresponding parabolas of  $\Delta-m_h(\alpha+v)^2/2$. The latter energy defines the absorption edge.
In all cases, $\Pi_{\eta_e\eta_{\text{h}}}<0$ below the corresponding absorption edge. This implies that the effective interaction is attractive. 

Using explicit expressions for $\Pi'_{\eta_e\eta_{\text{h}}}$ from Table~\ref{table:pis}, we find  that the bound state energies in the corresponding symmetry channels are determined by the following equations
\begin{subequations}
    \bea
    \frac{1}{u}&=&\frac{2}{\sqrt{y/\beta^2-1}}\tan^{-1}\frac{1}{\sqrt{y/\beta^2-1}}+\ln\frac{L}{y},\;\text{for the }\;
    J_z=0^-\;\text{channel};\label{eigA2}\\
    \frac{1}{u}&=&\frac{2}{\sqrt{y-1}}\tan^{-1}\frac{1}{\sqrt{y-1}}+\ln\frac{L}{y},\;\text{for the }\;
    J_z=0^+\;\text{channel};\label{eigA1}\\
    \frac{1}{u}&=&\frac{1}{\sqrt{y-1}}\tan^{-1}\frac{1}{\sqrt{y-1}}+\frac{1}{\sqrt{y/\beta^2-1}}\tan^{-1}\frac{1}{\sqrt{y/\beta^2-1}}+\ln\frac{L}{y},
    \;\text{for the}\;
    J_z=\pm 1\;\text{channel}.\label{eigE}
    \eea
\end{subequations}
Here $u=m\lambda/2\pi\hbar^2$ is the dimensionless coupling constant, $\beta=(\alpha-v)/(\alpha+v)$,  $y=2 (\Delta-\epsilon)/m_h(\alpha+v)^{2}$, and $L=\Lambda/2m(\alpha+v)$ is the dimensionless cutoff energy. The right-hand sides of the above equations exhibit 1D-like, square-root singularities near the corresponding absorption edges, i.e., at $y\to \beta^2$ for the $
J_z=0^-$ channel
and at $y\to 1$ for the $
J_z=0^+$ and $
J_z=\pm 1$ channels. This guarantees that bound states exist even for infinitesimally weak interaction. At weak coupling ($u\ll 1$), one can keep only the most singular terms in the right-hand sides of Eqs.~(\ref{eigA1}-\ref{eigE}).  The simplified equations are then solved analytically and give
\begin{subequations}
    \bea
    \epsilon_{
        0^-}&=&E_g^{
        0^-}-\frac{\pi^2u^2}{2} m_h(\alpha-v)^2 ,\\
    \epsilon_{
        0^+}&=&E_g^{
        0^+}-\frac{\pi^2u^2}{2} m_h(\alpha+v)^2,\\
    \epsilon_{
        \pm 1}&=&E_g^{
        \pm 1}-\frac{\pi^2u^2}{8} m_h(\alpha+v)^2.
    \eea
\end{subequations}
It is clear that $\epsilon_{
    0^+}\leq \epsilon_{
    \pm 1}<\epsilon_{
    0^-}$. 
A numerical solution of Eqs.~(\ref{eigA2}-\ref{eigE}) [cf. Fig.~\ref{Fig_ExDep} in MT] shows that this hierarchy of the bound states remains the same even beyond weak coupling. If the relative signs of $\alpha$ and $v$ are chosen to be opposite, i.e., $\alpha v<0$, the $0^+$ and $0^-$ eigenenergies are interchanged.

As explained in MT, the 1D-like threshold singularities in Eqs.~(\ref{eigA2}-\ref{eigE}) are due to the fact that the density of states of the Rashba spectrum exhibits a square-root van Hove singularity at the bottom of the band \cite{chaplik:2006,cappelluti:2007,zak:2010,berg:2012}. 

\subsection{B.~Explicit solution of the Schroedinger equation}\label{sec:Sch}
In this section, we construct an explicit solution of the Schroedinger equation corresponding to the Hamiltonian (Eq.~(2) of MT). 
As before,
we focus on an exciton with total momentum $\vec{k}=0$ and consider a model case of point-like attraction between electron and hole, $V(\vec{r})=-\lambda \delta(\vec{r})$. The advantage of an explicit solution is in that it allows one to classify the eigenstates found in the previous section by the eigenvalues of the total angular momentum.

In the momentum-space representation, the corresponding Schroedinger equation reads
\bea
\mathcal{H}(\vec p)\psi(\vec p)=-\lambda
\int \frac{d^2p}{(2\pi\hbar)^2}\psi(\vec p)
\label{Sch}
\eea
where 
\bea
\mathcal{H}(\vec p)=-\left[\left(\frac{\vec p^2}{2m_{\text{h}}}+\Delta-\epsilon\right)\mathbb{1}_{\sigma_{\text{e}}}\otimes\mathbb{1}_{\sigma_{\text{h}}}+v(\boldsymbol{\sigma}_{\text{e}}\times\vec p)\cdot\hat{\vec{z}}\otimes\mathbb{1}_{\sigma_{\text{h}}}-\alpha\mathbb{1}_{\sigma_{\text{e}}}\otimes(\boldsymbol{\sigma}_{\text{h}}\times\vec p)\cdot\hat{\vec{z}}\right].
\eea 
Indeed, in matrix form we have
\begin{equation*}
\begin{pmatrix}
\frac{\vec p^2}{2m_{\text{h}}}+\Delta-\epsilon & -i\alpha pe^{-i\phi} & ivpe^{-i\phi}&0\\
i\alpha pe^{i\phi}&\frac{\vec p^2}{2m_{\text{h}}}+\Delta-\epsilon & 0 & ivpe^{-i\phi}\\
-ivpe^{i\phi} & 0&\frac{\vec p^2}{2m_{\text{h}}}+\Delta-\epsilon & -i\alpha pe^{-i\phi}\\
0&-ivpe^{i\phi}&i\alpha pe^{i\phi}&\frac{\vec p^2}{2m_{\text{h}}}+\Delta-\epsilon
\end{pmatrix}
\begin{pmatrix}
e^{im\phi}\psi_1(p)\\
e^{i(m+1)\phi}\psi_2(p)\\
e^{i(m+1)\phi}\psi_3(p)\\
e^{i(m+2)\phi}\psi_4(p)
\end{pmatrix}
=
2\pi\lambda\begin{pmatrix}
\bar{\psi}_1\delta_{m,0}\\
\bar{\psi}_2\delta_{m,-1}\\
\bar{\psi}_3\delta_{m,-1}\\
\bar{\psi}_4\delta_{m,-2}
\end{pmatrix},
\end{equation*}
where $\bar{\psi}_i\equiv \int \frac{dp\,p}{(2\pi)^2}\psi_i(p)$.
It can be readily seen that the $\phi$-dependent terms cancel out leaving a set of of integral equations for $\psi_1(p)\dots\psi_4(p)$.

Writing a formal solution of Eq.~(\ref{Sch}) as
\bea
\psi(\vec p)=-\lambda \mathcal{H}^{-1}(\vec p)\int \frac{d^2p'}{(2\pi\hbar)^2}\psi(\vec p'),
\eea
we close the system by integrating the equation above over $\vec p$. This yields a set of linear equations for amplitudes of the wave function averaged over the momentum
\bea
\left[1+\lambda \int \frac{d^2p}{(2\pi\hbar)^2}\mathcal{H}^{-1}(\vec p)\right]\int \frac{d^2p'}{(2\pi\hbar)^2}\psi(\vec p')=0.
\eea
In a matrix form, the equation above reads
\bea
\begin{pmatrix}
    \frac{N}{2}&0&0&0\\
    0&\frac{N}{2}&-M&0\\
    0&-M&\frac{N}{2}&0\\
    0&0&0&\frac{N}{2}
\end{pmatrix}\begin{pmatrix}
    \bar{\psi}_1\delta_{m,0}\\
    \bar{\psi}_2\delta_{m,-1}\\
    \bar{\psi}_3\delta_{m,-1}\\
    \bar{\psi}_4\delta_{m,-2}
\end{pmatrix}=0,
\eea
where $M$ and $N$ are the same as in Eq.~ (\ref{evalues}).
Hence, the $m=0,-2$ states are degenerate with eigenvalue equation $
N=0$, while the two $m=-1$ states have the eigenvalue equations $N\pm 2M=0$. 
These eigenvalue equations  match exactly the ones given in Eq.~(\ref{explicit_evalues}), obtained in the scattering-matrix formalism.

Since the operator of the total angular momentum $\hat J_z=
\mathbb{1}_{\sigma_{\text{e}}}\otimes\mathbb{1}_{\sigma_{\text{h}}}(-i\partial_\phi)+
\frac{1}{2} \mathbb{1}_{\sigma_{\text{e}}}\otimes(\sigma_{\text{h}})_z+\frac {1}{2} (\sigma_{\text{e}})_z \otimes\mathbb{1}_{\sigma_{\text{h}}}
$ commutes with the Hamiltonian,
the eigenstates of the latter can be classified by the eigenvalues of $\hat J_z$. 
A simple calculation shows that  $J_z=m+1$ for our spinor.
Hence, $J_z=1,0,-1$ for $m=0,-1,-2$, respectively. This confirms the classification of the eigenstates adopted in MT and the previous section of SI.

\subsection{C.~Bound state energy for Coulomb potential}
The 1D nature of the Rashba spectrum also helps to generalize the results presented above for the case of Coulomb interaction, which is  a more realistic model than a delta-function one. Since the effective Hamiltonian of the two-body problem, Eq.~(2) of MT, is the of the Rashba type, one can follow Ref.~\cite{chaplik:2006} which analyses a bound state formed around an attractive impurity in a Rashba electron system. In our case, electrons and holes are located on the surface of a semi-infinite insulator with the dielectric constant $\varepsilon_{\text{i}}$, hence the effective dielectric constant for the Coulomb problem 
is $\varepsilon_{\text{eff}}=(\varepsilon_{\text{i}}+1)/2$. At frequencies relevant to exciton formation ($\lesssim 1$\, eV)~
$\varepsilon_{\text{i}}
\sim\varepsilon_\infty\sim
25$ \cite{Reijnders2014PRB}, while the Bohr velocity $v_B=e^2/\varepsilon_{\text{eff}}\approx 1.1$\,eV\AA, which is substantially smaller than the effective Rashba velocity in the $E_{1,2}$ channel $\alpha+ v=7.2$\,eV\AA. Therefore, the Coulomb interaction can be considered as a perturbation to the free Rashba Hamiltonian, and the effective 1D nature of the problem allows one to use the well-known result for a weakly bound
state in 1D: $E_b=-\left[\int dx U(x)\right]^2 m/2$, where $U(x)$ is potential of attractive perturbation \cite{Landau:77}. For $U(x)\propto 1/x$, the logarithmic divergence of the integral $x\to\infty$ 
can be regularized either by invoking  screening by free carriers \cite{chaplik:2006}
or by effectively resumming the perturbation theory \cite{grimaldi:2008}. The detailed analysis of numerical parameters that the second scenario is
more relevant to Bi$_2$Se$_3$.
In this case, the variational estimate for 
the ground state energy 
is given by \cite{grimaldi:2008}
\bea
\epsilon_0=\Delta-4\text{Ry}^*\ln^2
\frac{2 m_h(\alpha+v)a_{B}}
{\text{e}^{2}\hbar}
,\label{coulomb}
\eea
where $\text{Ry}^*=m_he^4/2\varepsilon^2_{\text{eff}}$ is the effective Rydberg energy,
$a_B=\hbar^2\varepsilon_{\text{eff}}/m_he^2$ is the effective Bohr radius, and $\text{e}=2.718\dots$ is the base of the natural logarithm.
A $\ln^2$ enhancement of the bound state energy is similar to that for a hydrogen atom in the strong magnetic field \cite{Landau:77}. As in the latter case, the excited states follow the usual (3D) Balmer sequence: $\epsilon_n=\Delta-\text{Ry}^*/n^2$, with $n=1,2\dots$. 
A better estimate for the ground state energy can be obtained from the numerical solution of the Schroedinger equation presented 
in Fig.~2 of Ref.~\cite{grimaldi:2008}. With parameters for Bi$_2$Se$_3$, this solution gives $0.22$\,eV for the bound state energy. Given that the experimentally observed threshold energy is $2.5$\,eV and the photoluminescence peak is at $2.3$\,eV, the binding energy of the exciton is $0.2$\,eV, which is very close to the theoretical value.
\begin{table*}
    \caption{\label{table:pis}Real and imaginary parts of $\Pi_{\eta_e,\eta_{\text{h}}}$ [Eq.~(\ref{eq: bubble})]. Here, 
        $\beta=(\alpha-v)/(\alpha+v)>0$, $y=2 (\Delta-\epsilon)/m_h(\alpha+v)^{2}$, $L=\Lambda/2m(\alpha+v)$.}
    \begin{adjustwidth}{-0cm}{}
        {\renewcommand{\arraystretch}{1.5}
            \renewcommand{\tabcolsep}{0.2cm}
            \begin{tabular}{|c|p{4.2cm}|p{5.45cm}|p{4.4cm}|}
                \hline
                \quad & $0<y<1$	& $y>1$ & $y<0$\\
                \hline
                $-\frac{2\pi\hbar^2}{m}\Pi'_{++}$ & \begin{eqnarray*}
                    && \frac{1}{\sqrt{1-y}}\ln\frac{1-\sqrt{1-y}}{1+\sqrt{1-y}}
                    \\&+&
                    \ln\frac{L}{y}
                \end{eqnarray*} & \begin{eqnarray*}
                    &&\frac{2}{\sqrt{y-1}}\left(\tan^{-1}\frac{1}{\sqrt{y-1}}-\frac{\pi}{2}\right)\nn\\&+&\ln\frac{L}{y}
                \end{eqnarray*} &\begin{eqnarray*}
                    &&\frac{1}{\sqrt{1+|y|}}\ln\frac{\sqrt{1+|y|}-1}{\sqrt{1+|y|}+1}\nn\\
                    &+&\ln\frac{L}{|y|}	
                \end{eqnarray*} \\ \hline
                $-\frac{2\pi\hbar^2}{m}\Pi''_{++}$ & \begin{eqnarray*}
                    0
                \end{eqnarray*} & \begin{eqnarray*}
                    0
                \end{eqnarray*} & \begin{eqnarray*}
                    1-\frac{1}{\sqrt{1+|y|}}
                \end{eqnarray*}\\
                \hline
                $-\frac{2\pi\hbar^2}{m}\Pi'_{--}$ & \begin{eqnarray*}
                    &&\frac{1}{\sqrt{1-y}}\ln\frac{1-\sqrt{1-y}}{1+\sqrt{1-y}}\nn&+&\ln\frac{L}{y}
                \end{eqnarray*} & \begin{eqnarray*}
                    &&\frac{2}{\sqrt{y-1}}\left(\tan^{-1}\frac{1}{\sqrt{y-1}}+\frac{\pi}{2}\right)\nn
                    &+&\ln\frac{L}{y}		\end{eqnarray*} & \begin{eqnarray*}
                    &&\frac{1}{\sqrt{1+|y|}}\ln\frac{\sqrt{1+|y|}-1}{\sqrt{1+|y|}+1}\nn&+&\ln\frac{L}{|y|}
                \end{eqnarray*}\\ \hline
                $-\frac{2\pi\hbar^2}{m}\Pi''_{--}$ & \begin{eqnarray*}
                    \frac{2}{\sqrt{1-y}}
                \end{eqnarray*} & \begin{eqnarray*}
                    0
                \end{eqnarray*} & \begin{eqnarray*}
                    1+\frac{1}{\sqrt{1+|y|}}
                \end{eqnarray*}\\
                \hline
                \quad & $0<y<\beta^{2}$	& $y>\beta^{2}$ & $y<0$\\
                \hline
                $-\frac{2\pi\hbar^2}{m}\Pi'_{-+}$ & \begin{eqnarray*}
                    &&\frac{1}{\sqrt{1-\frac{y}{\beta^{2}}}}\ln\frac{1-\sqrt{1-\frac{y}{\beta^{2}}}}{1+\sqrt{1-\frac{y}{\beta^{2}}}}\nn &+&\ln\frac{L}{y}
                \end{eqnarray*} & \begin{eqnarray*}
                    &&\frac{2}{\sqrt{\frac{y}{\beta^2}-1}}\left(\tan^{-1}\frac{1}{\sqrt{\frac{y}{\beta^2}-1}}-\frac{\pi}{2}\right)\nn\\&+&\ln\frac{L}{y}\nn
                \end{eqnarray*} & \begin{eqnarray*}
                    &&\frac{1}{\sqrt{1+\frac{|y|}{\beta^2}}}\ln\frac{\sqrt{1+\frac{|y|}{\beta^2}}-1}{ \sqrt{\frac{|y|}{\beta^2}+1}+1}\nn&+&\ln\frac{L}{|y|}
                \end{eqnarray*}\\ \hline
                $-\frac{2\pi\hbar^2}{m}\Pi''_{-+}$ & \begin{eqnarray*}
                    0
                \end{eqnarray*} & \begin{eqnarray*}
                    0
                \end{eqnarray*} & \begin{eqnarray*}
                    1-\frac{1}{\sqrt{1+\frac{|y|}{\beta^2}}}
                \end{eqnarray*}\\
                
                \hline
                $-\frac{2\pi\hbar^2}{m}\Pi'_{+-}$ & \begin{eqnarray*}
                    &&\frac{1}{\sqrt{1-\frac{y}{\beta^{2}}}}\ln\frac{1-\sqrt{1-\frac{y}{\beta^{2}}}}{1+\sqrt{1-\frac{y}{\beta^{2}}}}\nn&+&\ln\frac{L}{y}
                \end{eqnarray*} & \begin{eqnarray*}
                    &&\frac{2}{\sqrt{\frac{y}{\beta^2}-1}}\left(\tan^{-1}\frac{1}{\sqrt{\frac{y}{\beta^2}-1}}+\frac{\pi}{2}\right)\nn&+&\ln\frac{L}{y}
                \end{eqnarray*} & \begin{eqnarray*}
                    &&\frac{1}{\sqrt{1+\frac{|y|}{\beta^2}}}\ln\frac{\sqrt{1+\frac{|y|}{\beta^2}}-1}{\sqrt{1+\frac{|y|}{\beta^2}}+1}\nn&+&\ln\frac{L}{|y|}
                \end{eqnarray*}\\
                \hline
                $-\frac{2\pi\hbar^2}{m}\Pi''_{+-}$ & \begin{eqnarray*}
                    \frac{2}{\sqrt{1-\frac{y}{\beta^2}}}
                \end{eqnarray*} & \begin{eqnarray*}
                    0
                \end{eqnarray*} & \begin{eqnarray*}
                    1+\frac{1}{\sqrt{1+\frac{|y|}{\beta^2}}}
                \end{eqnarray*}\\
                \hline
        \end{tabular}}
    \end{adjustwidth}
\end{table*}

\clearpage

\section{2.~Electronic band structure of B\lowercase{i}$_2$S\lowercase{e}$_3$\label{SM:band}}
\begin{figure}[t]
    \centering
    \includegraphics[width=5cm]{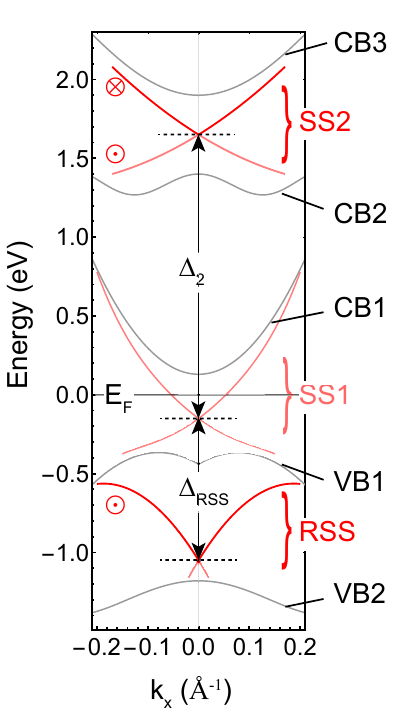}
    \caption{\label{FigSM_band}
        The electronic band structure near the Brillouin zone center as illustrated in Fig.~\ref{Fig1}(a) in the MT.
        The red lines denote the surface states, whereas the gray lines denote bulk bands.
        The in-plane spin orientation of the surface state electrons are denoted by $\odot$ and $\otimes$.
    }
\end{figure}

In this section, we describe the details of the Bi$_2$Se$_3$ band structure, and construction of Fig.~\ref{Fig1}(a) of the MT.
The Fermi level in the single crystal from which all the data in the MT were collected is about 0.15\,eV above the Dirac point of SS1, 
as
determined by STM 
on
crystals 
from
the same batch \cite{Dai2016}.
The dispersions of the Dirac surface states near the Fermi surface (SS1) are obtained 
by fitting 
the 
ARPES 
data 
of Nomura et al.~\cite{Nomura2014}
into 
Eq.\,(4) in Ref.~\cite{Fu2009}:
\begin{equation}
E_{SS1}(p)=-0.15+\frac{p^2}{2m^\ast} \pm\sqrt{v_1^2p^2+\lambda^2p^6\cos^2 3\theta}
\end{equation}
Fitting 
yields $m^\ast\approx 0.066$\,eV$^{-1}$\AA$^{-2}$ for the effective mass,
$v_1\approx 2.4$\,eV{\AA} for the velocity,  $\lambda\approx 50$\,eV\AA$^3$ for the warping 
coefficient.

Dispersions of the unoccupied Dirac surface states (SS2)
are 
obtained by fitting 
the ARPES 
data
of Sobota et al.~\cite{Sobota2013} 
into the following equation
\begin{equation} \label{eq:linear}
E_{SS2}(p)=\Delta_{SS2}+\frac{p^2}{2m_e}+v p,
\end{equation}
with the zero of energy set at the SS1 Dirac point.
Such a fit gives
$\Delta_{SS2}\approx 1.8$\,eV, 
$v
\approx 2.0$\,eV{\AA}, and 
$m_e\approx 0.16$\,eV$^{-1}$\AA$^{-2}$.

Dispersions of the
Rashba surface states (RSS) are 
obtained by fitting 
the ARPES data 
of Soifer et al.~\cite{Soifer2017} 
into the following equation
\begin{equation} \label{eq:Rashba}
E_{RSS}(p)=\Delta_{RSS}+
|\alpha|
p
-\frac{p^2}{2m_{\text{h}}}.
\end{equation}
Here, $\Delta_{RSS}\approx 0.9$\,eV,
the Rashba coefficient $|\alpha|\approx 5.2$\,eV{\AA}, and the effective mass $m_h\approx 0.036$\,eV$^{-1}$\AA$^{-2}$.

In Eq.~(1) of the MT, $\Delta=\Delta_{SS2}+\Delta_{RSS}\approx 2.7$\,eV
and $m_e$ is neglected compared to $m_h$.
Dispersions of 
the
bulk conduction band (CB$_1$) and 
of the 
valence band (VB1) are extracted from Ref.~\cite{Bianchi2010}; dispersions of the conduction bands CB$_2$ and CB$_3$ are extracted from Ref.~\cite{Sobota2013}; dispersion of the valence band VB$_2$ is extracted from Ref.~\cite{Soifer2017}. 
All of the
above dispersions are obtained by fitting 
the data from the corresponding references in 
phenomenologically chosen
polynomial functions.

\begin{figure}[t]
    \centering
    \includegraphics[width=17cm]{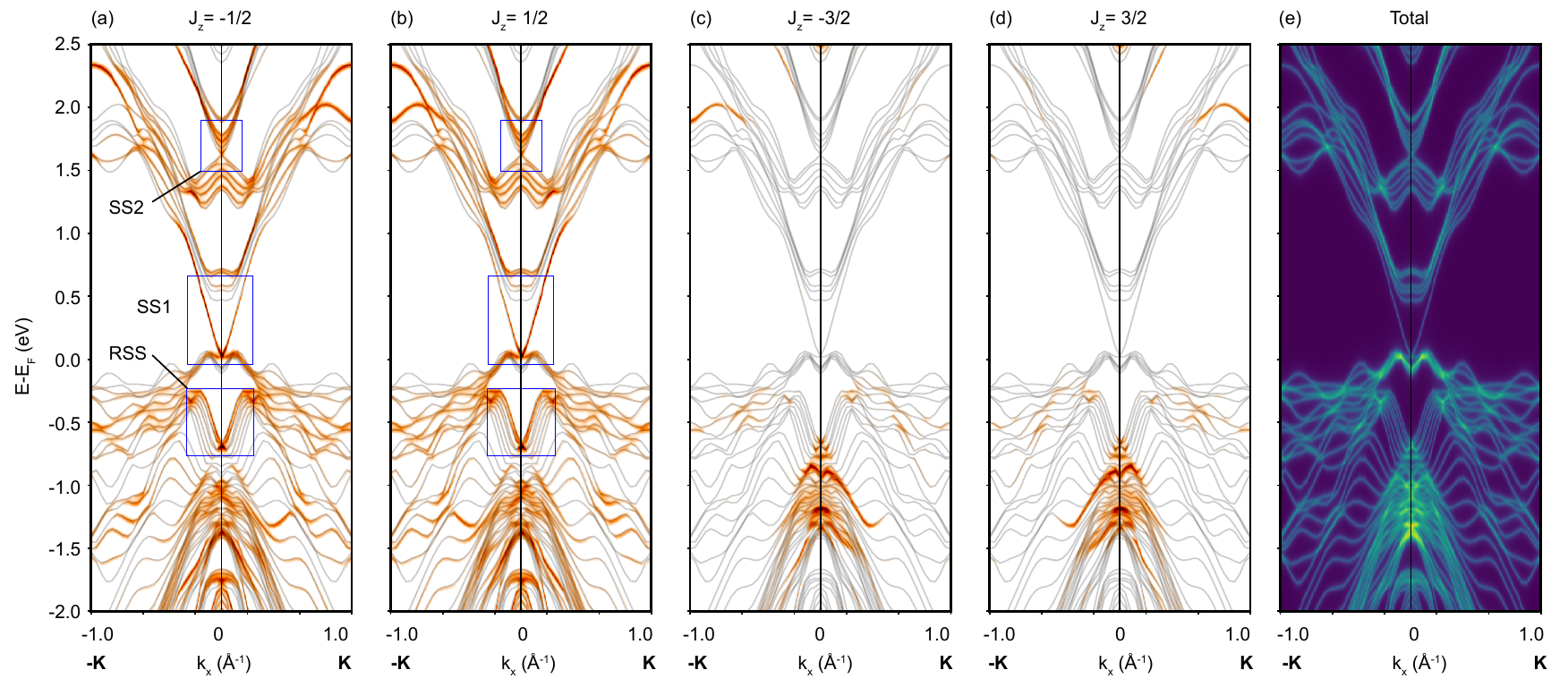}
    \caption{\label{FigSM:DFT}
        Calculated band structures, projected onto the top quintuple layer for (a) $J_z=-1/2$, (b) $J_z=1/2$, (c) $J_z=-3/2$, (d) $J_z=3/2$ states, and (e) without considering the angular momentum.
        The blue squares in (a) and (b) highlight the surface states in Fig.~\ref{FigSM_band}, SS1, SS2 and RSS.
        All figures are shown along the $\Gamma$--K cut in the Brillouin zone of the hexagonal lattice.
    }
\end{figure}

In Fig.~\ref{FigSM:DFT} we show the calculated band structure projected onto the top quintuple layer (QL), and plot along the $\Gamma$--K cut in the Brillouin zone of the hexagonal lattice.
The blue squares in (a) and (b) highlights the surface states in Fig.~\ref{FigSM_band}, SS1, SS2 and RSS.

The band structure calculations were performed using the Quantum Espresso software package~\cite{QESPRESSO2009,QESPRESSO2017}, using fully relativistic PBEsol exchange functionals~\cite{Perdew2008PRL}. 
The wavefunction and density energy cutoffs were chosen to be 40\,Ry and 400\,Ry, respectively, and we used a $15\times 15\times 1$ momentum grid.
The system was arranged in a slab geometry with a 32\AA vacuum layer, relaxed to obtain equilibrium coordinates, but keeping the inner 2 layers fixed at bulk coordinates.
After computing the band structure, the wave functions were projected onto the eigenstates of total angular momentum for each ion. 
Fig.~\ref{FigSM:DFT}(e) shows the projections onto the top QL without considering the angular momentum, where the projections onto the $J_z=1/2$ states, including both $J=3/2$ and $J=1/2$ manifolds, are shown in (a)--(b), and the projections onto the $J_z=1/2$ states are shown in (c)--(d).

The calculation shows that the surface band structure in Bi$_2$Se$_3$ are dominated by $J_z=\pm 1/2$ states.
In particular, the SS2 and upper-cone of RSS states in Fig.~\ref{FigSM_band} relevant to this paper are well reproduced by $J_z=\pm 1/2$ projections in Fig.~\ref{FigSM:DFT}(a)--(b), consistent with the assumptions in Eq.(1) of MT.

\section{3.~Sample and temperature dependence of polarized photoluminescence}\label{SM:sample}
\begin{figure}[t]
    \centering
    \includegraphics[width=7cm]{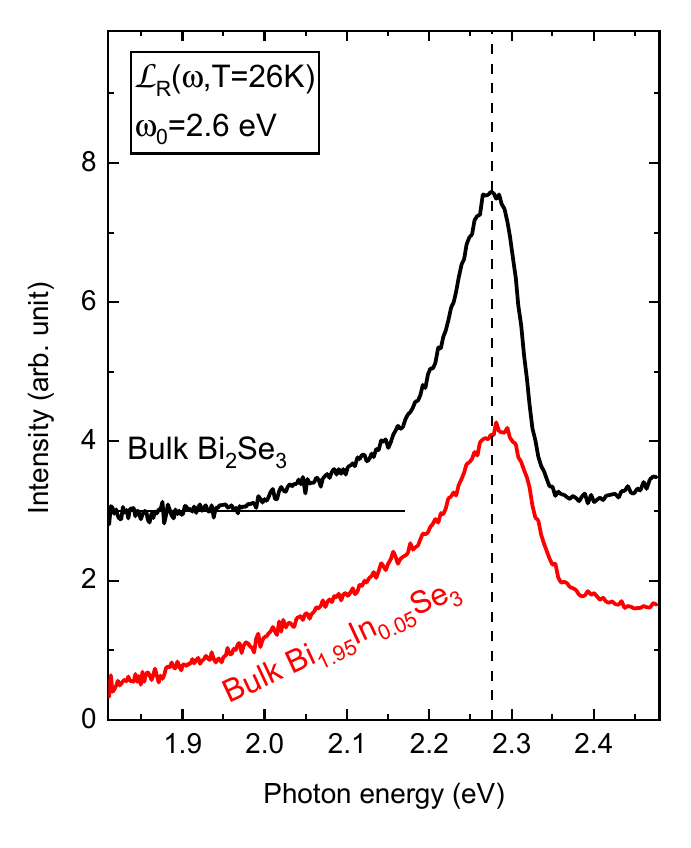}
    \caption{\label{FigSM_SampleDep}
        $\mathcal{L}_{R}(\omega,T)$ is plotted against photon energy for pristine and indium substituted Bi$_2$Se$_3$ samples, measured with 2.6\,eV right-CP excitation at 26\,K.
        Spectra are shifted vertically for clarity.
        Dashed line is guide to the eye indicating the PL peak center of the bulk Bi$_2$Se$_3$ single crystal.
    }
\end{figure}

Figure~\ref{FigSM_SampleDep} compares the polarized PL from bulk Bi$_2$Se$_3$ and Bi$_{1.95}$In$_{0.05}$Se$_3$ single crystals.
Both samples show qualitatively the same line shape and intensity.
The PL peak in bulk Bi$_{1.95}$In$_{0.05}$Se$_3$ crystal shows slight broadening compared to the pristine Bi$_2$Se$_3$ crystals, which could be due to increased inhomogeneity level from indium substitution.
The indium substituted samples are likely possessing higher carrier concentration.
However, the similar spectral line shapes demonstrate that the PL peak is a robust feature insensitive to the Fermi level.

\begin{figure}[t]
    \centering
    \includegraphics[width=7cm]{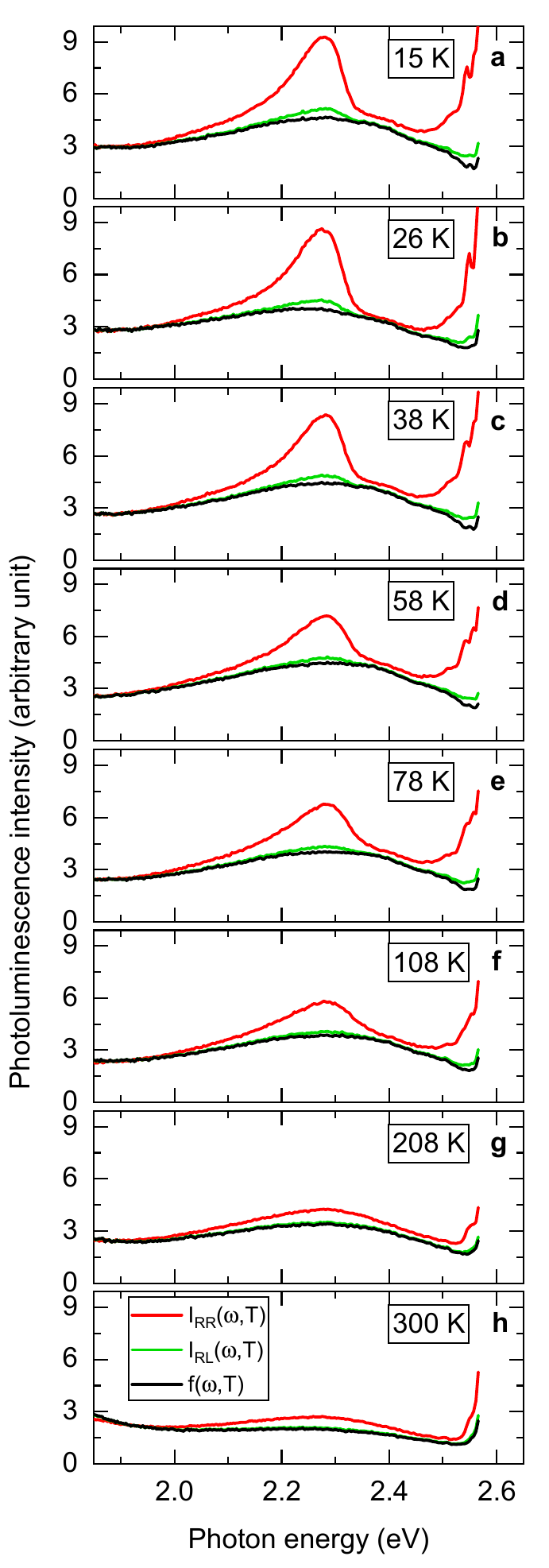}
    \caption{\label{FigSM_TDep}
        The right- and left-handed PL spectra, $I_\text{RR}(\omega,T)$ and $I_\text{RL}(\omega,T)$, measured with 2.60\,eV right-CP excitation on bulk single crystal between 15 to 300\,K.
        The black curves show the unpolarized background, \( f(\omega,T)=\frac{I_\text{RL}(\omega,T)-r(T)\cdot I_\text{RR}(\omega,T)}{1-r(T)} \), with the depolarization ratio $r(T)=0.1$ for all temperatures shown here.
    }
\end{figure}

Figure~\ref{FigSM_TDep} plots the right- and left-handed circularly polarized PL spectra, $I_\text{RR}(\omega,T)$ and $I_\text{RL}(\omega,T)$, measured with 2.60\,eV (476\,nm) right-CP excitation on bulk single crystals between 15 to 300\,K.
The black curves in Fig.~\ref{FigSM_TDep} are the unpolarized PL background, $f(\omega,T)$, which we assume to be featureless around 2.3\,eV.
By defining an energy independent depolarization ratio \( r(T)\equiv \frac{I_\text{RL}(\omega,T)-f(\omega,T)}{I_\text{RR}(\omega,T)-f(\omega,T)} \), 
we can rewrite the unpolarized PL background as \( f(\omega,T)=\frac{I_\text{RL}(\omega,T)-r(T)\cdot I_\text{RR}(\omega,T)}{1-r(T)} \).
Then, $r(T)$ is determined by minimizing sharp features in $f(\omega,T)$ around the 2.3\,eV PL peak.
The determined $r(T)$ depend strongly on excitation, as shown in Fig.~1(e), but very weakly on temperature.
For 2.60\,eV excitation, $r(T)$ were found to be about 0.1 for all temperatures between 15 to 100\,K.
Above 100\,K, the error for determination of $r(T)$ is large due to broadening of the PL peak.
However, one can still clearly see the polarized PL from the raw data (Fig.~\ref{FigSM_TDep}), and no significant depolarization is observed up to room temperature.

\clearpage

%
%
\end{document}